\documentclass[fleqn,usenatbib]{mnras}

\usepackage{newtxtext,newtxmath}
\usepackage[T1]{fontenc}

\DeclareRobustCommand{\VAN}[3]{#2}
\let\VANthebibliography\thebibliography
\def\thebibliography{\DeclareRobustCommand{\VAN}[3]{##3}\VANthebibliography}

\usepackage[utf8]{inputenc}
\usepackage{natbib}
\usepackage{amsmath}
\usepackage{graphicx}
\usepackage{txfonts}
\usepackage{hyperref}
\usepackage{verbatim}
\usepackage{xspace}
\usepackage{color}
\usepackage{breqn}
\usepackage{multicol,lipsum}
\usepackage{etoolbox}
\usepackage{enumitem}
\usepackage{xcolor}
\usepackage{caption}
\usepackage{subcaption}

\usepackage{threeparttable}
\usepackage{booktabs}
\usepackage{ulem} %% for sout
\usepackage{dirtytalk}

\newcommand\ipole{{\tt IPOLE}\xspace}
\newcommand\THEMIS{{\tt THEMIS}\xspace}
\newcommand\ehtim{{\tt eht-imaging}\xspace}

\title[Testing Bayesian inference of GRMHD model parameters from VLBI data]{Testing Bayesian inference of GRMHD model parameters from VLBI data}

\author[A. I. Yfantis, et al.]{
A. I. Yfantis$^{1}$ \thanks{a.yfantis@astro.ru.nl}, S. Zhao$^{2}$, R. Gold$^{3}$, M. Mo\'scibrodzka$^{1}$,
A. E. Broderick$^{4,5,6}$.
\\
$^{1}$Department of Astrophysics, IMAPP, Radboud University, 6500 GL Nijmegen, The Netherlands\\
$^{2}$Shanghai Astronomical Observatory, Chinese Academy of Sciences, 80 Nandan Road, Shanghai 200030, People’s Republic of China\\
$^{3}$CP3-Origins, University of Southern Denmark, Campusvej 55, 5230 Odense, Denmark\\
$^{4}$Perimeter Institute for Theoretical Physics, 31 Caroline Street North, Waterloo, ON, N2L 2Y5, Canada\\
$^{5}$Department of Physics and Astronomy, University of Waterloo, 200 University Avenue West, Waterloo, ON, N2L 3G1, Canada\\
$^{6}$Waterloo Centre for Astrophysics, University of Waterloo, Waterloo, ON, N2L 3G1, Canada\\
}
\date{Accepted XXX. Received YYY; in original form ZZZ}

\pubyear{2024}

\begin{document}
\label{firstpage}
\pagerange{\pageref{firstpage}--\pageref{lastpage}}
\maketitle

% Abstract of the paper
\begin{abstract}
Recent observations by the Event Horizon Telescope (EHT) of supermassive black holes M87* and Sgr A* offer valuable insights into their spacetime properties and astrophysical conditions. Utilizing a library of model images ($\sim 2$ million for Sgr A*) generated from general-relativistic magnetohydrodynamic (GRMHD) simulations, limited and coarse insights on key parameters such as black hole spin, magnetic flux, inclination angle, and electron temperature were gained. The image orientation and black hole mass estimates were obtained via a scoring and an approximate rescaling procedure. Lifting such approximations, probing the space of parameters continuously, and extending the parameter space of theoretical models is both desirable and computationally prohibitive with existing methods. To address this, we introduce a new Bayesian scheme that adaptively explores the parameter space of ray-traced, GRMHD models. The general relativistic radiative transfer code \ipole is integrated with the EHT parameter estimation tool \THEMIS. The pipeline produces a ray-traced model image from GRMHD data, computes predictions for VLBI observables from the image for a specific VLBI array configuration and compares to data thereby sampling the likelihood surface via an MCMC scheme. At this stage we focus on four parameters: accretion rate, electron thermodynamics, inclination, and source position angle. Our scheme faithfully recovers parameters from simulated VLBI data and accommodates time-variability via an inflated error budget. We highlight the impact of intrinsic variability on model fitting approaches. This work facilitates more informed inferences from GRMHD simulations and enables expansion of the model parameter space in a statistically robust and computationally efficient manner.

\end{abstract}

% Select between one and six entries from the list of approved keywords.
% Don't make up new ones.
\begin{keywords}
Accretion, accretion disks -- Black hole physics --  Methods: data analysis --  Methods: statistical --  Techniques: high angular resolution --  (Galaxies:) quasars: supermassive black holes
\end{keywords}

%%%%%%%%%%%%%%%%%%%%%%%%%%%%%%%%%%%%%%%%%%%%%%%%%%

%%%%%%%%%%%%%%%%% BODY OF PAPER %%%%%%%%%%%%%%%%%%

\section{Introduction}

The Event Horizon Telescope (EHT) is a millimeter Very Long Baseline Interferometric (mm-VLBI) array capable of resolving compact (sizes of $\sim 20 \, \mu as$) event horizon scale structures around supermassive black holes in M87 (called M87*) (\citealt{Doeleman2012Sci1224768,eht2019a,eht2019b,eht2019c,eht2019d,eht2019e,eht2019f}) and in the Galactic Center (called Sagittarius A*, abbreviated Sgr~A*). (\citealt{Doeleman2008Nature07245,Fish2011ApJL727_2_L36,Johnson2015Sci.acc7087,Lu:2018,eht2022a,eht2022b,eht2022c,eht2022d,eht2022e,eht2022f}).

As recently demonstrated in \citet{eht2019e,eht2019f,eht2022e,eht2022f}, EHT also enables inferences of the astrophysical conditions present in the relativistic environment in the immediate vicinity of a black hole horizon by comparing the VLBI data to theoretical models that predict the on-sky emission map. For this purpose substantial libraries of diverse source models, model comparison techniques, and parameter estimation tools are being built and constantly improved.

In particular, general relativistic magnetohydrodynamics (GRMHD) numerical models of inefficiently radiating accretion flows onto a black hole combined with general relativistic radiative transfer (GRRT) models predict the appearance of the two EHT main targets  (\citealt{moscibrodzka:2009,dexter:2009,moscibrodzka2014,moscibrodzka2016,gold2017,alejandra:2018,chael:2019}). 
These numerical models of magnetized accretion flows depend on a few key physical parameters such as: (i) the spin of the black hole, (ii) magnetic flux threading the horizon (iii) the mass accretion rate onto the black hole, (iv) the electron thermodynamics (here simply modeled via $R_{high}$, and (v) the orientation of the system with respect to the observer. 
Constraining these free parameters of GRMHD simulations via EHT observations can give us quantitative estimates of black hole mass and spin, insights into how gravitational energy is converted into radiation in strong gravity and what mechanism launches the astrophysical jets such as the one observed in M87* (e.g., \citealt{hada:2013,kim:2018}).  

To constrain physical parameters of M87* and SgrA*, \citet{eht2019e,eht2022e} created a {\it static} library of approximately $60,000$ GRMHD model images for M87* and about $1,800,000$ for Sgr~A* and then compared the libraries to the EHT data via various scoring procedures  \citep{eht2019e,eht2019f,eht2022e,eht2022f}. 
 The two main ones are: a) average image scoring (AIS, used for total intensity data), where snapshots from simulations are turned into synthetic VLBI observations and compared to real data; and b) snapshot characterization (used mostly for polarimetric images), where a suite of image properties such as, e.g. resolved degree of linear and circular polarizations and their directions are compared to polarimetric characteristics of the reconstructed source images. 

In the existing scoring procedure of the EHT to total intensity data (AIS scoring), a) only the total flux of the image (for a fixed accretion rate), b) the mass of the black hole, and c) the position angle of the model are estimated in adaptive fashion over the entire possible range. In EHT terminology this is called snapshot scoring, and creates distributions using all snapshots available (see previous par.). After that, for each model (combination of GRMHD+GRRT parameters), consisting of $ \sim500$ snapshots an average image is created. This averaged image generates synthetic data that are compared with the real data, given a standard deviation from the spread of the snapshots. Then, given a passing criterion (e.g. cumulative distribution $\in[2.5\%,97.5\%]$) the models pass or fail the AIS test (see \citet{eht2019e}). In this procedure parameters such as inclination angle, electron heating parameter $R_{\rm high}$, black hole spin and magnetic flux on the horizon are sampled sparsely in the limited range. Moreover, the two latter parameters are fixed for a given GRMHD simulation and changing them is computationally expensive as it requires running an entire GRMHD simulation. 

%\textcolor{red}{In this work we propose and test a pipeline for continuously/adaptively fitting the less computationally expensive parameters, such as the accretion rate (that will change the total flux of the model image), viewing angle and the electron heating parameters which is achievable under the present computational capabilities. This is particularly useful for the black hole in the center of our Galaxy where, e.g., our viewing angle of the system is weakly constrained but the black hole mass is known precisely.}  

 A key aspect of accretion flows and GRMHD simulations, variability, is particularly challenging for inference pipelines. Variability refers to the inhomogeneity of an accretion flow in a spatial and temporal sense. These two aspects are often intertwined, since a spatial variability (a disk without azimuthal symmetry for example) is magnified by temporal variability, where the directions and particularities of this asymmetrical flow are changing direction and even structure over time. 
 %I remove this sentece because it is not true. 
 %Many GRMHD simulations (mostly magnetically arrested disks) are able to reproduce this effect, but the magnitude is over-produced by a factor of two \citep{eht2022d}, and the structural changes cannot be robustly tested with current observations. 
 This means that when comparing EHT data with simulations it is necessary to provide many snapshots of a simulation to test if any of them resembles the source, and even then it will be an approximation. Hence the inference pipelines need to be capable of matching two images that are a priori different. In AIS this is done by the usage of the aforementioned large model libraries.
 
In this paper, we propose a new Bayesian parameter estimation procedure by integrating the GRRT code {\tt ipole} \citep{Monika2018} with EHT/VLBI data analysis framework THEMIS (\cite{THEMIS-CODE-PAPER} and Sect. \ref{subsec:fitting} in this paper) and enable the {\it adaptive} GRRT parameter estimation given an arbitrary GRMHD snapshot. In the improved parameter estimation scheme, the parameters defined in GRRT, e.g. the inclination angle, the accretion rate and the plasma thermodynamics parameter will be {\it adaptively} sampled across the entire parameter space to compute the posterior distributions via Bayesian inference. Notice that in this approach the large amount of memory for statically storing the image library is not necessary. We also show that our pipeline could provide a robust framework to account for the variability challenges, see Section~\ref{subsec:variability2}. Additionally, the pipeline is designed to be highly parallelized and extensible, which is the important first step towards the large scale computation of {\it adaptive} parameter estimation from GRMHD simulations in the future. 

To assess the accuracy and efficiency of our new parameter extraction scheme, we first pick an arbitrary set of GRRT model parameters and generate an image from a GRMHD simulation. 
Next, we simulate the EHT 2017 observation by assuming the above image has the same celestial position, mass and distance as Sgr~A* (but the procedure can be also adopted for M87*) and generate synthetic mm-VLBI data, including visibility amplitudes and closure phases.
%this is already mentioned in the next para or even used in the current paper
%\footnote{The fitting can be extended to closure amplitudes and even polarimetric data.}. 
Since, for our fitting routine we decide to use only the closure phases, the interstellar media scattering effect is not considered at this primary step (which is important and complicated for fitting archive Sgr A* data, see e.g. (\citet{johnson2018, issaoun2019,Issaoun2021ApJ915.99}). Additionally, we apply standard thermal noise (\citealt{Chael2016ApJ829.11C,Chael2018ApJ857.23}) and systematic errors ($1\%$, $10\%$ or $30\%$ depending on the case).
Then we use a Markov chain Monte Carlo (MCMC) algorithm to sample the posteriors of the parameters from the underlying unknown distribution of all the physically possible models by comparing the synthetic data with the GRMHD+GRRT model. 
We perform well-controlled tests with a known \say{truth} value first by fitting two parameters and then extend it to fit four parameters simultaneously. Such an incremental approach provides clarity when interpreting the results.

% Lastly, as a first illustration we use the same pipeline to fit a single GRMHD model directly to an archival EHT 2009 observations of Sgr~A* and demonstrate four parameter estimation: the inclination angle, the position angle, the mass accretion rate onto the black hole and a electron thermodynamical parameter. 

In the future, the pipeline can be improved for instance with more realistic observational corruptions of model images (e.g., \citealt{blecher2017},  Janssen et al. 2019) or polarimetric models \citep{eht2021h}. \THEMIS as well as \ipole can handle different observing frequencies. Therefore, the pipeline presented here can naturally handle model fitting to upcoming EHT, ngEHT, and non-EHT data sets, e.g. to longer wavelengths VLBI observations of the EHT targets or AGN sources \citep{kim:2018,issaoun2019,eht2021MWL}.

The paper is organized as follows. In Sect.~\ref{sec:pipeline}, we describe the pipeline which produces the GRMHD+GRRT models of Sgr~A* (or M87*) at millimeter waves (Sect.~\ref{subsec:descipt_model}), the process of generating synthetic mm-VLBI observation data sets from the image (Sect.~\ref{subsec:descipt_observable}), the sampling methods (Sect.~\ref{subsec:fitting}). 
In Sect.~\ref{sec:test_synthetic}, we use the synthetic data (generated in Sect.~\ref{subsec:descipt_observable}) to test the adaptive parameter estimation pipeline by two parameter fitting (Sect.~\ref{subsubsec:two_fit}) as well as multi-parameter fitting (Sect.~\ref{subsec:multi_fit}). In Sect. \ref{subsec:variability1} we introduce time variability in the fitting algorithm, and in \ref{subsec:variability2} we propose a few ways to account for it.
We summarize the results and conclude in Sect.~\ref{sec:discussion}.

%%%%%%%%%%%%%%%%%%%%%%%%%%%%%%%%%%%%%%%%%%%%%%%%%%
\section{Parameter Estimation Pipeline: Description}\label{sec:pipeline}
%%%%%%%%%%%%%%%%%%%%%%%%%%%%%%%%%%%%%%%%%%%%%%%%%%

\subsection{Physical model and  model parameters}\label{subsec:descipt_model}

\begin{figure*}
  \includegraphics[width=0.97\linewidth,trim={0cm 0.8cm 0 0},clip]{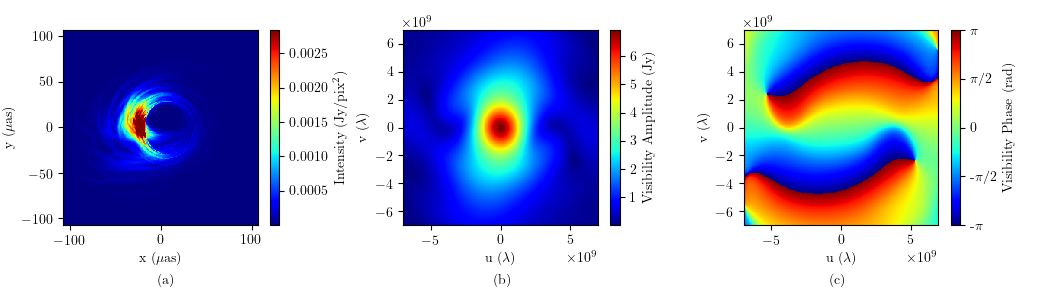}\\
\includegraphics[scale=0.376,trim={0 0cm 0cm 0cm},clip]{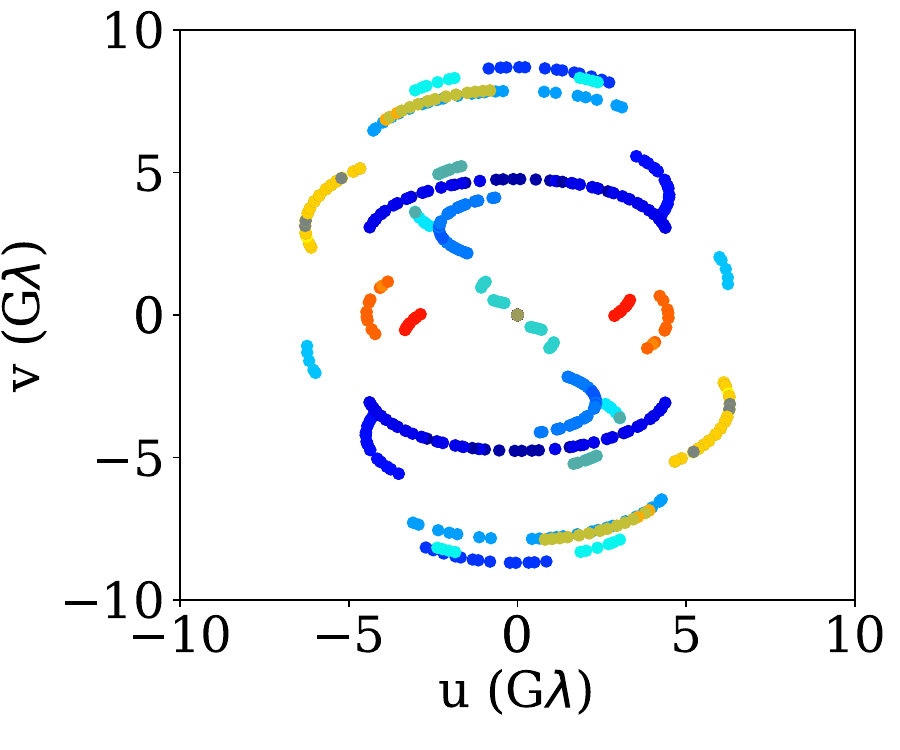}
         \includegraphics[scale=0.376,trim={0cm 0cm 0 0cm},clip]{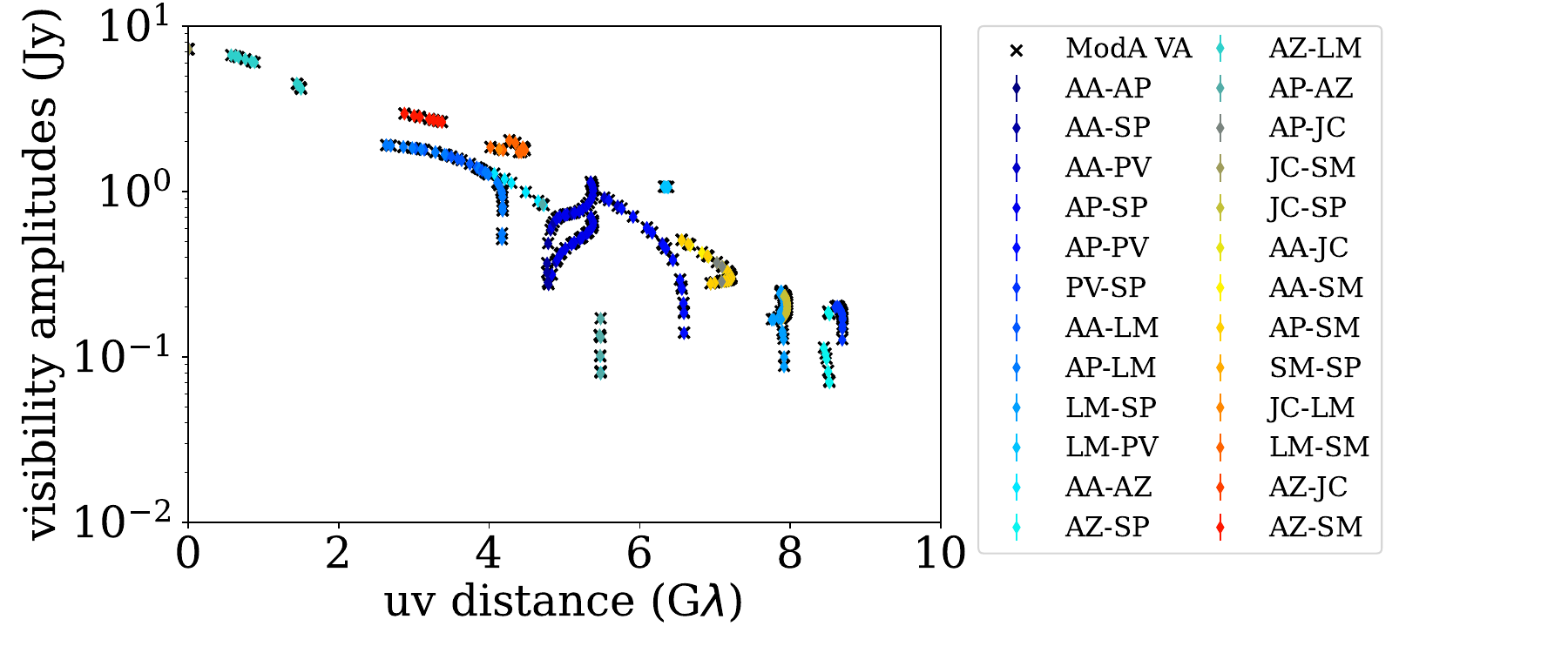}
         \includegraphics[scale=0.4,trim={0cm 0cm 0 0cm},clip]{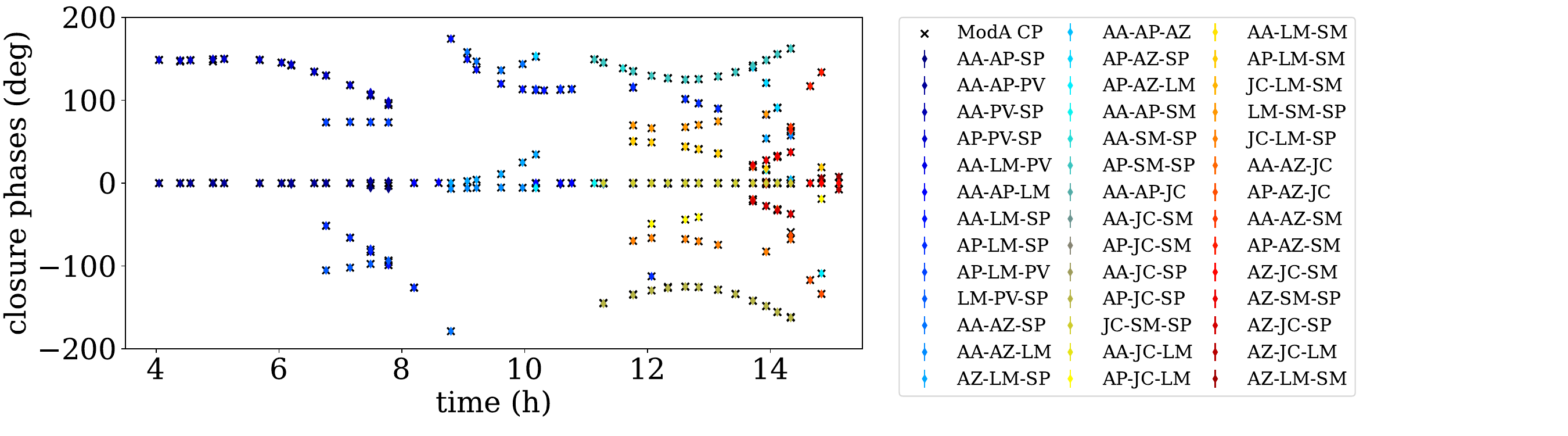}
        \caption{
        {\bf Top panels:} 230\,GHz image of GRMHD model of Sgr~A*. Physical parameters used to generate image are: $M_{\rm unit}=3\times 10^{18} grams$, $R_{\rm high}=3$, $R_{\rm low}=3$, $i=60^{\circ}$, $PA=0^{\circ}$. 
  Here, the model image has high resolution of $256 \times 256$ pixels. The color shows the emission intensity (Stokes $\mathcal{I}$). The middle and right panels show the amplitudes and phases of the complex visibility function which is generated by the Fourier transformation (FT) of the model image. {\bf Bottom panels:} Synthetic EHT\,2017 data generated using \ehtim \, and \THEMIS. The middle left panel shows the baseline $(u,v)$ coverage of the observation and the middle right panels displays the visibility amplitudes as a function of (u,v) distance. The colors code different baselines. The bottom panel shows the synthetic closure phases. The different colors refer to different EHT station triangles. Black crosses denote the data modeled within \THEMIS which are in excellent agreement with those from \ehtim.}
        \label{fig:synthetic_data}
\end{figure*}

Extracting physical parameters from EHT observations requires 
a model for the accretion flow onto a compact object and a model for the arising emission. 

Our model describes an accreting black hole within ideal-GRMHD simulation and is therefore intrinsically dynamical. The simulation starts with a torus of plasma in Keplerian, equatorial orbit around a Kerr black hole \citep{fishbone1976} that would be in hydrostatic equilibrium in absence of magnetic fields. The torus is then seeded with weak magnetic fields and
the evolution of such configuration is computed by solving the equations of ideal general relativistic magnetohydrodynamics \citep{gammie:2003}. Simulations show that magnetic turbulence is developed, which acts as an effective source of viscosity \cite{Hawley1991,Balbus1998}, thereby causing matter to accrete onto the black hole. In this process some material escapes the system in the form of winds and jets that may or may not be visible in the model image depending on radiation properties and electron thermodynamics.

The free parameters of the GRMHD-informed model images can be divided into two categories: numerical parameters and physical parameters. The numerical parameters are e.g., numerical grid size and resolution of the simulation. Physical parameters include: black hole spin parameter $a_*\in(0,1)$\footnote{The black hole spin is usually given in dimensionless units where $a_*=0$ describes Schwarzschild black hole and $a_*=1$ is maximally rotating Kerr black hole} or the normalized magnetic flux threading the horizon in the relaxed, steady state, which is the fundamental parameter describing the state of the magnetic field along the standard and normal evolution (SANE) and magnetically arrested disk (MAD) regimes \citep{sasha:2011,porth:2019}. In the present paper, we utilize time-slices of an existing 3D GRMHD SANE simulation of \cite{shiokawa2013} (applied to model Sgr A* in \citealt{moscibrodzka2014}), with the black hole spin parameter $a_*=0.9375$ and the adiabatic index of $13/9$. We pick snapshots where the turbulence is fully developed and the simulation exhibits roughly a steady state behavior in which at least the interior regions have not retained their initial conditions.

Next, we generate synthetic images (intensity maps) from this simulation snapshot using the ray-tracing \& radiative transfer code \ipole~\footnote{\url{https://github.com/moscibrodzka/ipole}} \citep{Monika2018}, which was tested against other radiative transfer codes used in the original \ipole paper as well as in \citet{Gold2020} and \citet{Prather:2023}.The fast-light approach is used throughout this work for both synthetic data and Bayesian runs. The main radiative processes considered in computing images from the GRMHD simulation is synchrotron emission and synchrotron self-absorption. 
Generating images of a particular astrophysical source requires rescaling the dimensionless GRMHD simulations from geometrized unit system (G=M=c=1) to c.g.s. units. The scaling requires providing the mass of the central black hole which will also set the length scale of the system according to $GM_{\rm BH}/c^2 \mathrm{[cm]}$ and time scale units $GM_{\rm BH}/c^3 \mathrm{[s]}$. The scaling also requires providing the mass unit parameter $M_{\rm unit}$ that scales the density of the plasma around the black hole, i.e., the density of the matter in the accretion flow is $\rho_{\rm c.g.s.} = \rho_{\rm code} M_{\rm unit}/{\mathcal L}^3$ (Notice that $M_{\rm unit}$ also scales the accretion rate onto the black hole and strength of magnetic field at the same time, see \citealt{Monika2009} for details). It is therefore in principle necessary to redo the ray tracing / radiative transfer computation whenever these parameters are varied, which is what we pursue here in contrast to an approximate scaling approximation designed to avoid the additional computational cost. The free parameters used to model the differences between electron and proton temperatures in various regions of different magnetization are $R_{\rm low}$ and $R_{\rm high}$ (motivated by \citealt{moscibrodzka2016} and \citealt{ressler2015}, see also \citealt{eht2019e}).
Specifically, the proton to electron temperature ratio reads:
\begin{equation}\label{eq:tpte}
\frac{T_p}{T_e}=R_{\rm low} \frac{1}{1+\beta^2} + R_{\rm high}\frac{\beta^2}{1+\beta^2}.
\end{equation}
Given $T_{\rm p}$, which is equal to the gas temperature in the GRMHD simulation, we can compute $T_{\rm e}$
using Eq.~\ref{eq:tpte} and the synchrotron emissivities thus depend on the assumed $R_{\rm high},R_{\rm low}$ parameters. As $R_{\rm high}$ and $R_{\rm low}$ are not two independent parameters, $R_{\rm low}$ is usually fixed to be a constant ($R_{\rm low}=1$ in \citealt{eht2019e}).
The inclination angle, $i$, and position angle, $PA$ (orientation of the image on the sky with respect to the celestial North pole), are two remaining parameters of the model describing the geometrical orientation of the system with respect to the observer's line of sight. Finally, the distance of the observer to the source has to be assumed.

For current models only black hole spin and magnetic flux require an independent GRMHD simulation. This leads to higher computational efficiencies especially for higher dimensions and the ability to estimate the posterior distribution for each parameter including black hole mass and spin. We also drop the assumption about optical depth made in \cite{eht2019e,eht2022e} where a crude flux rescaling was applied, treating BH mass and total flux as scale-free parameters (to within a limited range). Such a scaling can at most be valid in a finite range and more specifically for matter that is sufficiently optically thin. Our method reperforms the ray-tracing and radiative transfer on  every likelihood evaluation and hence drops these assumptions.
%Instead here we redo the radiative transfer properly/independently on every step, including $M_{\rm unit}$ in the fitting parameters. 
This is in itself a significant improvement over the current method.

In this work we focus on parameter estimation by scaling the dimensionless GRMHD simulations to  Sgr~A* system but other black hole masses can be assumed. We therefore fix the mass of the black hole to $M_{\rm BH,Sgr~A*}= 4.1 \times 10^6 {\rm M_{\odot}}$ and distance $D_{\rm Sgr A*}=8.5 \mathrm{kpc}$ \citep{g:2018}. Other model parameters 
$i$, $PA$, $M_{\rm unit}$, $R_{\rm low}$, $R_{\rm high}$ are allowed to float. This list can easily be generalized to include any parameter in the ray-tracing and radiative transfer code used.

In Fig.\ref{fig:synthetic_data} (left-most upper panel) we show an example of an arbitrarily chosen sets of parameters of appearance of the 3D GRMHD simulation scaled to Sgr~A* system parameters as seen by an observer on Earth. 
The model image is generated at an observational wavelength of $\lambda=1.3\mathrm{mm}$ ($\nu=230 \mathrm{GHz}$) at which EHT currently operates (in the future EHT will also operate at $0.87\mathrm{mm}/345 \mathrm{GHz}$).  
%Notice, that at $\lambda \gtrsim 1$mm, Sgr~A* structure will be additionally washed out due to scattering of radio waves by free electrons in the Milky Way spiral arms \citep{johnson2018} which is not included in this figure and neither in our fitting routines. 

\subsection{VLBI data products and synthetic data generation}\label{subsec:descipt_observable}

EHT is an interferometer which detects the sparsely sampled Fourier 
components of the image of the source on the sky, called visibilities.
The visibility $V(u,v)$ is the 2-dimensional Fourier transformed complex function of intensity distribution $I(x,y)$ defined by  e.g., \citet{TMS} as:
\begin{equation}
V(u,v)=\iint{ I(x,y)e^{-2\pi i(ux+vy)} } dxdy.
\end{equation}
The visibility is by definition a complex function with amplitude $A$ and phase $\phi$: $V(u,v)=A e^{-i \phi}$. 
In Fig.~\ref{fig:synthetic_data} (middle and right-most upper panels) we show the visibility amplitude and phase computed based on the GRMHD model image.

%In case of Sgr~A* (but not M87*), the visibility amplitudes should be additionally modified to include the mentioned smearing effects caused by scattering of radio waves by free electrons in the galaxy. In the present work we do not use the visibility amplitudes, so there has not been a need for scattering effects. 

The visibility amplitudes are subject of a future work, so they are not discussed here. 
In the present we utilize closure phases, the sum of the complex visibility phases along a closed triangle baseline, which is:
\begin{equation}
\label{eq:CP}
    \Phi_{i,j,k}=\arg{(V_{ij}V_{jk}V_{ki})},
\end{equation}
where $V_{ij}$,$V_{jk}$ and $V_{ki}$ are the visibility of baseline ij, jk and ki. 
Due to degeneracy amongst possible triangles, an array with $N$ antennas has $(N-1)(N-2)/2$ {\it independent} closure phases (\citealt{TMS}, \citealt{Blackburn_2020}). For the 2017 EHT observations $N=8$ which gives 21 independent closure phases but in general one can form up to 56 closure phase triangles ($N(N-1)(N-2)/3!$) assuming that the source is visible at all sites during an observing window.
%\ar{are you sure? This seems to be the number of baselines.. For the triangles I would expect to be the number of combinations of 3 from a set of N minus some triangles that can't not observing simultaneously.}
%\ar{sorry, it just seemed a bit complicated to do discuss it here. but pls, go ahead.}
In practice, using 2017 EHT $(u,v)$ coverage, we generate 41 closure phases. The simulated observation is roughly 11 hours long. The main advantage of the closure quantities is that they are 
insensitive to station-based errors
%comment out details because station based errors is broad enough to encapsulate all the complexity of the telescopes
%, including atmospheric errors or station gains
%are eliminated in closure phases
\citep{Chael2018ApJ857.23}. 
%commented out because this is mentioned in the next para at the end
%When modeling closure phases one may consider taking into account thermal noise (see Appendix~\ref{app:errors} for discussion of data errors).

In Fig.~\ref{fig:synthetic_data} (middle and bottom panels) we show
the $(u,v)$ coverage, synthetic visibility amplitudes and closure phases.
Our example of synthetic VLBI data is generated based on GRMHD image assuming the following parameters: $i=60^{\circ}$, $M_{\rm unit}=3\times 10^{18} grams$, $R_{\rm high}=3$, $R_{\rm low}=3$, and position angle $PA=0^{\circ}$. For the following $M_{\rm unit}$ will always be in $(grams)$. The image has 128$\times$128 pixels (see Appendix~\ref{app:image_res} for discussion of image resolution), and we assume that the source is on the celestial sphere where the ascension and the declination are same as Sgr~A*. 
The synthetic VLBI data are generated using the \ehtim~library\footnote{\url{https://github.com/achael/eht-imaging}} \citep{Chael2016ApJ829.11C,Chael2018ApJ857.23}. We simulate the EHT observation with EHT 2017 array configuration to observe Sgr~A* with the baseline $(u,v)$ coverage matching the EHT observation on April $7^{th}$ 2017 \citep{eht2022b}. The center of the observational frequency band is $229.1\mathrm{GHz}$ and the bandwidth is $1.8\mathrm{GHz}$. 

Our synthetic EHT data are time-averaged along $(u,v)$ tracks into 10 minutes scans. Furthermore, the data have been treated to account for typical noises, such as thermal and systematic noises usually considered when analysing EHT data (see, e.g., \citealt{eht2019d}, \citealt{eht2024a,eht2024b}). The thermal noise (tn) is to account for fluctuations in the telescope during "observation", while the systematic errors(syser), set at $1\%$, $10\%$ or $30\%$, capture additional uncertainties from the instrument.
%ensure Gaussian character and add confidence in the truth of the measurements.  

When fitting model to synthetic or real EHT data,
the calculation of visibility amplitudes and closure phases from the given model image is the nested part of EHT data analyze framework \THEMIS\footnote{\url{https://github.com/PerimeterInstitute/Themis}}.
\THEMIS \ is a massively-parallel, modular, flexible and extensible framework, containing all the utilities necessary to compare EHT data to a variety of model predictions for these data sets, including visibility amplitudes, closure phases and more. The FFTW3\footnote{FFTW is a publically available C subroutine library for computing the discrete Fourier transform in one or more dimensions, of arbitrary input size, and of both real and complex data. \url{http://www.fftw.org/}} is the fast Fourier transformation (FFT) library used to transform the intensity distribution on the celestial plane to the visibility of each uv data sets, which are read from input data files. 
In Fig.\ref{fig:synthetic_data} we also show that the visibility amplitudes and the closure phases calculated with THEMIS match perfectly those produced by \ehtim~library.

% \begin{figure}
%          \includegraphics[width=0.94\linewidth,trim={0 0cm 0cm 0cm},clip]{figures/simobs_all.pdf}
%         \caption{The simulated EHT observation of visibility amplitudes (top) and closure phases (bottom) for different cases: 1) without any additional corruptions (blue), 2) with scattering screen added (green), 3) with scattering and thermal noise added (red), 4) with scattering, thermal noise and systematic noise added (black).}
%         \label{fig:synthetic_data_detail}
% \end{figure}

\subsection{Model fitting}
\label{subsec:fitting}
\subsubsection{Likelihood and priors}\label{subsec:descipt_lklhd_prior}
%\RG{[I will rephrase this section.]}

\THEMIS  \, carries out Bayesian parameter estimation via MCMC sampling the log-likelihood.
In the Bayesian statistics, if given the prior probability distribution of parameters to estimate, the posterior probability distribution is constrained by the likelihood function. 
The likelihood and the prior are defined by the user. The log-likelihood of closure phases is
\begin{equation}
    \label{eq:LKLHD_CP}
    \mathcal{L}(\vec{p})=-\sum_j\frac{\Delta^2\left(\Phi_j-\hat{\Phi}_j(\vec{p})\right)}{2\sigma_j^2},
\end{equation}
where $\vec{p}$ is the vector of parameters to estimate, $\Phi_j$ and $\hat{\Phi}_j(\vec{p})$ are the observed and modeled closure phases, $\Delta(x)$ is the angular difference in the range $[-180^\circ, 180^\circ)$ and $\sigma_j = \sqrt{\text{tn}^2+\text{syser}^2*\hat{\Phi}_j(\vec{p})^2}$.
A link to the more traditional approach is the relation $\mathcal{L}=-\chi^2/2$, from where it follows that

\begin{equation}
    \label{eq:x_sq}
    \chi^2(\vec{p})=\sum_j\frac{\Delta^2\left(\Phi_j-\hat{\Phi}_j(\vec{p})\right)}{\sigma_j^2}\,\, ; \,\, \chi_{\rm eff}^2 = \frac{\chi^2}{n_{d}-n_{f}}, 
\end{equation}
where $n_d = 358 $ is the number of data points, affected by the data character, visibility amplitudes (VA) or closure phases (CP), and the observation specifics, number of telescopes, time, etc. The number of freedom is $n_f = 2$ for two-parameter fit and 4 for the multi parameter fit, equal to the number of parameters being fitted simultaneously.

In this likelihood definition, we assume that the errors in the closure phases have Gaussian distribution. However when signal-to-noise ratio (SNR) is low, the error distribution is more likely to be non-Gaussian \citep{TMS}. How the error distribution and SNR affect the fitting accuracy is discussed in \cite{THEMIS-CODE-PAPER}. Another problem of this likelihood is that it is unable to treat fitting multi-epoch data. Both, the non-Gaussian errors and the multi-epoch observation fitting are beyond the scope of the present work.

Similarly to the phases, the log-likelihood for visibility amplitudes can be calculated as
\begin{equation}
    \label{eq:LKLHD_VA}
    \mathcal{L}(\vec{p})=-\sum_j\frac{\left[|V|_j-\hat{|V|}_j(\vec{p})\right]^2}{2\sigma_j^2},
\end{equation}
We adopt prior for each parameter separately. Due to the lack of the knowledge of the true model parameters and for keeping the approach as agnostic as possible, we adopt flat prior for all parameters. It is worth mentioning that flat prior is not equivalent to non-informative prior, but it is sufficient for this fit. 

\begin{table*}
   \caption{Physical and numerical model parameters explored in this work.}
    \centering
    \begin{threeparttable}
      \begin{tabular*}{\textwidth}{ccl}\toprule
         \midrule 
         \multicolumn{3}{c}{Physical Model Parameters}\\
         \midrule
         Name & Value/Range & Description\\
         \midrule
         $a_*$ & 0.9375 & Black hole spin parameter of a give GRMHD simulation snapshot; Fixed in this occasion. \\
         $i$ & ($0,180^\circ$) & Viewing angle (inclination) of the observer: $i=0^\circ$ is face-on, $i=90^\circ$ is edge-on.  \\
         $PA$ & [$-\pi,\pi$] &Position angle of the black hole spin on the sky, measured east of north. \\
         $M_{\rm unit}$ & [$10^{17},6\times10^{19}$] & Mass unit parameter scales the density of the plasma, hence ${\mathcal M}_{\rm unit} \sim \dot{M}$ $(g/s)$.\\
         $R_{\rm high}$ & [1,90] &  Describes coupling of $T_{\rm e}$ w/ $T_{\rm p}$ in regions of weak magnetization (high plasma $\beta$ region). \\ 
         $R_{\rm low}$ & 3 & Describes coupling of $T_{\rm e}$ w/ $T_{\rm p}$ in regions of strong magnetization (low plasma $\beta$ region).\\\toprule \midrule
         \multicolumn{3}{c}{Fitting Numerical Parameters }\\\midrule
         \multicolumn{3}{c}{Two parameter fit}\\
         \midrule
         Name & Value & Description\\
         \midrule
          image res. & $128^2$ & The size of model image, unit in pixels.\\
          $N_{\rm W}$ & 8& The number of chains in MCMC sampler.\\
          $N_{\rm T}$ &10& The temperature level of the parallel tempering in MCMC sampler. \\
            comm. freq.  & 2 & The interval MCMC steps between communication events of different tempered chains.\\
          burn-in &500 &The number of initial MCMC steps removed from chains when building posterior. \\
          end step & 5000 & The final MCMC step of the simulation \\\midrule
         \multicolumn{3}{c}{ Four parameter fit}\\
         \midrule
         image res. & $128^2$ & The size of model image, unit in pixels.\\
          $N_{\rm W}$ & 10& The number of chains in MCMC sampler.\\
          $N_{\rm T}$ &12& The temperature level of the parallel tempering in MCMC sampler. \\
            comm. freq.  & 2 & The interval MCMC steps between communication events of different tempered chains.\\
          burn-in &1000-3500 &The number of initial MCMC steps removed from chains when building posterior.\\
          end step & 10000 & The final MCMC step of the simulation \\
         \bottomrule \toprule
      \end{tabular*}
      \end{threeparttable}
 % }
   \label{tab:param}
\end{table*}

\subsubsection{Sampling parameters}\label{subsec:descript_sampling}

Markov chain Monte Carlo (MCMC) methods are frequently used to sample the posterior from prior with defined likelihood. 
In order to efficiently sample the underlying parameter space and to faithfully infer model parameters. Special care is needed on top of standard MCMC to avoid trapping in local extrema. We adopt a 
(parallel) tempering technique \citep{Swendsen1986PhRvL57.2607,Geyer1991}, in which high-tempered chains explore large regions in the parameter space with low precision while the low-tempered chains focus on small regions with high precision. The different tempered chains communicate and exchange their position information, which let the coldest chains (most accurate and used to be the final output) escape any local optimums. The scheme has been demonstrated to explore a variety of likelihood surfaces including multi-modal distributions and is well described in ~\cite{THEMIS-CODE-PAPER}. The sampler of choice for this project was the affine invariant (AI) method described in detail in \citet{Goodman2010CAMCoS5_65}. More possibilities and reasoning behind this choice can be found in Appendix~\ref{app:samplers}.

Within \THEMIS a variety of further options can be chosen, from VLBI data types, to sampling methods, number of walkers, temperatures, steps between communication of the walkers, number of processors per likelihood. 
Our final choice consists of fitting to closure phase data (CP), with the addition of thermal noise and systematic errors ($1\%,\,10\%,\,30\%$), the affine invariant sampler, and the number of walkers and temperatures: $N_{\rm W} = 8$, $N_{\rm T}=10$ for the two parameter fitting and $N_{\rm W} = 10$, $N_{\rm T}=12$ for the multi-parameter run (see next section). We have calculated the effective sampling size for our two models (A and B, see section \ref{sec:test_synthetic}) and we found minimum values of $\sim40$ and $\sim67$ respectively. We have not apply any thinning, i.e. retain only a subset of samples from the MCMC chains.

A summary of the physical and numerical parameters used in the calculations can be found in Table~\ref{tab:param}.

\section{Parameter Estimation Pipeline: Validation using Synthetic VLBI Data}\label{sec:test_synthetic}

%\subsection{Fitting setup}\monika{also this subsection seems like unnecessary repetitive rant}

%For our fiducial simulations we use a typical GRMHD+GRRT image (shown in the upper panels in Fig. \ref{fig:synthetic_data}, further referred to as "model A") to generate the synthetic data via simulating the EHT\,2017 observation. The physical parameters of model A are visible in Table \ref{tab:two-fit}.  

%In fitting test 2 we expand the analysis further in two ways, first by exploring a different error budget for the data of the truth, and second by creating an averaged image using 3 snapshots from the simulation, and using this image as the truth. These tests aim to probe an avenue to real data fitting, since we expect the real data to be smoothed out (like an average) and the first step to tackling the variability issue is simply inflating the errors of the data points to allow for more realisations of a particular snapshot.

In the following, first we perform two parameter fitting, using fitting parameter pairs of PA together with one of the other parameters (while keeping the remaining two fixed), the details of which are presented in section \ref{subsubsec:two_fit}. Then we raise both the number of $N_{\rm T}$ and $N_{\rm W}$ by 2 and perform an all parameter fitting simultaneously, which is presented in section \ref{subsec:multi_fit} and finally we discuss variability of the source and the data and how to tackle it in sections \ref{subsec:variability1}, \ref{subsec:variability2}

\subsection{Single snapshot, two parameter fitting} %$i$, $M_{\rm unit}$ or $R_{\rm high}$  with $PA$} 
\label{subsubsec:two_fit}

%\RG{[I need to understand again, what we did for the errors. We should use at least a typical thermal error and read that in consistently into the Themis data object (the sigma parameter passed in the Themis driver in the clospure phase data object). If this is done properly and we sampled well, we should be able to get a red chi^2 near 1. Maybe we can add a systematic error (multiplicative in visibitlites of 1\% mimicking non-closing errors from typical calibration uncertainties. ]}

In the first test, model A (shown in the upper panels in Fig. \ref{fig:synthetic_data}) is used to generate the truth synthetic data via simulating the EHT\,2017 observation.

The first pipeline test validates the scheme while fitting 2 parameters only: PA in combination with one of the 3 remaining parameters $i$, $M_{\rm unit}$ or $R_{\rm high}$. PA sampling alone is done more efficiently through an analytically marginalized likelihood however sampling $i$, $M_{\rm unit}$ and $R_{\rm high}$ requires repeating the GRRT simulation in every step. To initialize the MCMC chains we have chosen values in the middle of the range that we wish to explore (Table~\ref{tab:param}), using flat priors for all parameters.

\begin{figure*}
\centering
    \includegraphics[scale=0.643]{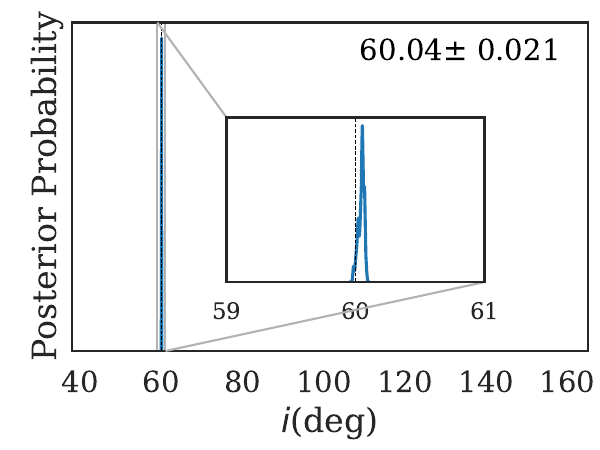}
    \includegraphics[scale=0.643]{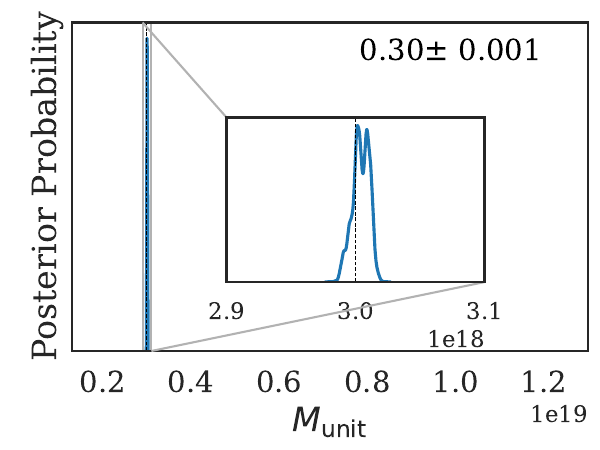}
    \includegraphics[scale=0.643]{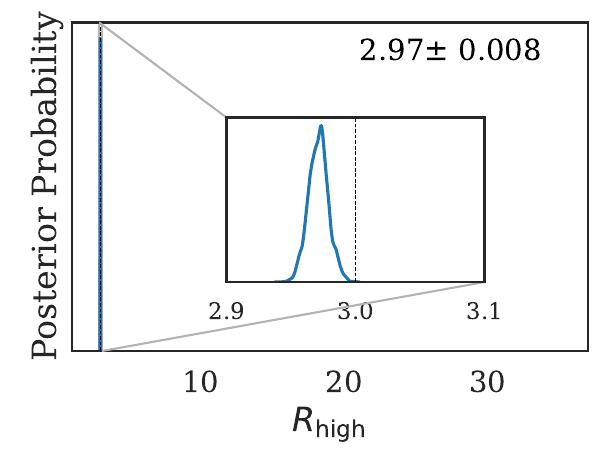}
    \includegraphics[scale=0.643]{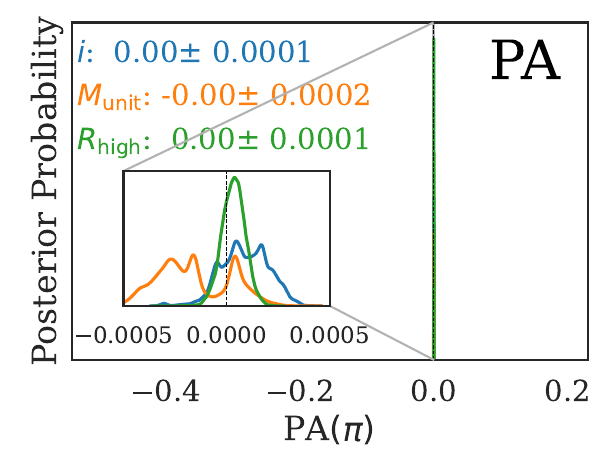}
    \includegraphics[scale=0.68]{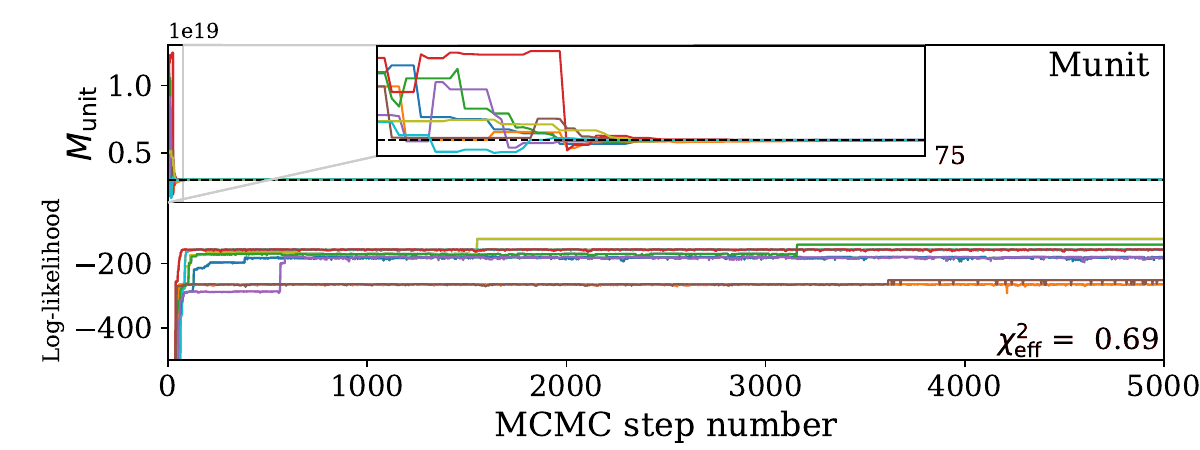}\\
    \caption{Results of the two parameters fits. In the left column panels show the posterior probability distribution. In the right column panels display the trace of the chains and log-likelihood of two parameter fitting for model A (with $1\%$ systematic errors). In the last row we plot the PA chains for all different two parameter fits together. The jobs ran for 5,000 MCMC and the burn-in for the distributions and the Gaussian estimates has been set to 500 MCMC. The $\chi^2_{\rm eff}$ reported in the right column panels are the smallest values achieved after the burn-in.}
    \label{fig:1P_fitting}
\end{figure*}

The test results are summarized in Fig.~\ref{fig:1P_fitting} where we show the posterior probability distribution (PD) for all parameters and the trace plot and the log-likelihood for $M_{\rm unit}$ as an example.
%The chains remain flat after the first 100 MCMC, with a good quality fit indicated by the smallest $\chi^2_{\rm eff}$ values, signature that the fitting is converging.
For all runs $\chi^2_{\rm eff} \le $ 1, as shown in Table~\ref{tab:two-fit}. We achieve effective sampling sizes larger than 600 for both parameters and a median split $\hat{R}=(0.9996,1.0017)$ in the two parameter fit.
%%%%%%%%%%%%%%%%%%%%%%%%%%%%%%%%%%%%%%%%%%%%%%%%%%%%%%%%%%%%%%%%%%%%%%%%%%%%%%%%%%%%
% \RG{[Based on what? How do you tell? What we show is that the chains are becoming flat to some extent that we should try to quantify.]} \ar{Is flatness and Chi values sufficient? } \RG{The flatness of the chains is basically the argument we are using, but other diagnostics could be computed. Send me the chains from the afs sampler and I should be able to show you more convergence diagnostics.}\ar{Sent in Slack} 
%%%%%%%%%%%%%%%%%%%%%%%%%%%%%%%%%%%%%%%%%%%%%%%%%%%%%%%%%%%%%%%%%%%%%%%%%%%%%%%%%%%%
\begin{table}
  \caption{Truth parameters and the pipeline performance for two parameter fitting, for a burn-in step of 500 MCMC. For each listed run, the PA and one other parameter is varied.}
  %\Shan{Is the table necessary?}
  %\RG{[Yeah point taken about redundancy, but I think it is nonetheless useful to show all inferred values next to their truth values in one place organized in a table.]}\ar{maybe now that xeff is not in plots, neither in text it makes more sense} }
    \centering
    \begin{tabular}{c|cccc}
         \hline\hline
         &\vspace{-0.8em}\\\vspace{0.4em}
         & truth & estimated &  $\chi_{\rm eff}^2$ 
         \\\hline\vspace{-0.8em}\\
         PA($\pi$)  &0&$0+O^{-4}$& - \\
         %\hline
         $i(^{\circ})$  &60&60.04$\pm0.021$&$0.66$\\
         $M_{\rm unit}(10^{18})$  &$3$&$3\pm0.001$&$0.69$\\
         $R_{\rm high}$ &$3$&$2.97\pm0.008$&$0.89$ \\
         \hline %\vspace{-0.8em}
         \hline
    \end{tabular}
      \label{tab:two-fit}
\end{table}

The posterior densities and the Gaussian fits of them have been made with a burn-in of 500 MCMC steps.
For all parameters except $R_{\rm high}$ the truth values are covered by the posterior (at 3 $\sigma$ in the case of inclination). By contrast, the PD of $R_{\rm high}$ has a shifted peak revealing a smaller than $1\%$ bias (from the median value) and misses the truth value altogether.

%%%%%%%%%%%%%%%%%%%%%%%%%%%%%%%%%%%%%%%%%%%%%%%%%%%%%%%%%

\begin{figure}
\centering
   \includegraphics[scale=0.68]{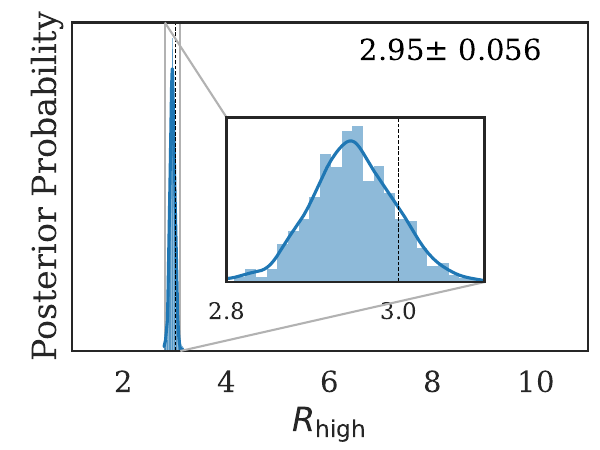}  
    \caption{The posterior probability distribution for two parameters ($R_{\rm high}$ and PA) fits when assuming systematic errors of $10\%$. Since we already have information about the truth this test is limited to 2,000 MCMC steps and a narrower prior for $R_{\rm high}\in(1,10)$. When fitting real VLBI data we would use a wider prior and run the pipeline longer.}
    \label{fig:Rh_10}
\end{figure}

To investigate $R_{\text{high}}$ offset we carry out several tests with three improvements, such as (i) adding more stations in the telescope array (from 2021, 2022), (ii) using CP and VA data, (iii) using a cut in the data where we only use CP points with S/N above 4 and lastly (iv) inflating the systematic errors at $10\%$. The posterior can be seen in Fig.~\ref{fig:Rh_10}.  Improvements (ii), (iii) and (iv) all resolve the parameter offset problem separately. This points to a bias caused by the closure phases, which is fixed either by the SNR cut or its effects are weakened when also VA data are included. The offset problem is also resolved when the systematic errors are inflated and the posterior covers a wider range of values.

Regarding the trace and log-likelihood plot (bottom panel in Fig. \ref{fig:1P_fitting}) the chains find the truth to within 1\% (even before 100 MCMC steps) from a wide parameter space. The log-likelihood values rise rapidly at first, and stabilize until the end. There is a mild spread in the log-likelihood values which is in line with the posterior distributions having two peaks. This was not a feature in $R_{\text{high}}$ for example. %\RG{[This is actually a sign that convergence may not be achieved. The sampler recently found a much better place. This implies that not much time has been spent on exploring this new better area yet. So run longer. We want a long boring flat line! That is a better indicator of convergence as discussed in the following sentence. BUT: The right thing to do is not to write it up but to push it to convergence and publish that.]} \ar{Changed phrasing -> Again.} 
Using Eq.~\ref{eq:x_sq} we have calculated the smallest $\chi^2_{\rm eff}$ for every parameter and the values are reported in Table~\ref{tab:two-fit}. 
%\RG{[If these values were correct our method would not work. Probably the error read in in Themis does not correspond to the error budget in the actual data? Maybe this is a tiny issue that disappears once we add a bit of noise.]} \ar{Note that the errors reported in Table 2 are a Gaussian approximation, using scipy.stats.bayes_mvs and could be a bad approximation.}

Two parameter fitting tests demonstrate
that it is possible to fit other parameters apart from PA, in an adaptive way, and that they converge to the truth values in a fast and stable fashion. The $\chi^2_{\rm eff}$ values for all runs below are close to 1 which further strengthen these claims. 
The two parameter fits for $R_{high}$ or $i$ already have advantage over the standard EHT image library which samples these variables rather sparsely (usually $R_{high}=1,10,40,160$ and $i=10,30,50...$).
%RG: Even in the simplified 2 parameter fit we can appreciate the advantage of sampling {\it any} values of $R_{high}$ or $i$ compared to the EHT image library.

\subsection{Single snapshot, four parameter fitting}\label{subsec:multi_fit}

Next, we
%use the affine invariant sampler, temperature level $N_{\rm T}=12$ and $N_{\rm W}=$10 chains to 
simultaneously sample four parameters (PA, $i$, $M_{\rm unit}$ and $R_{\rm high}$). We still fit only the CP data, using the same initial values and ranges of the parameters as in the previous two parameter fitting tests. 

\begin{figure*}
  \centering
  \includegraphics[scale=0.57,trim={0.5cm 0.5cm 0.5cm 0.1cm},clip] {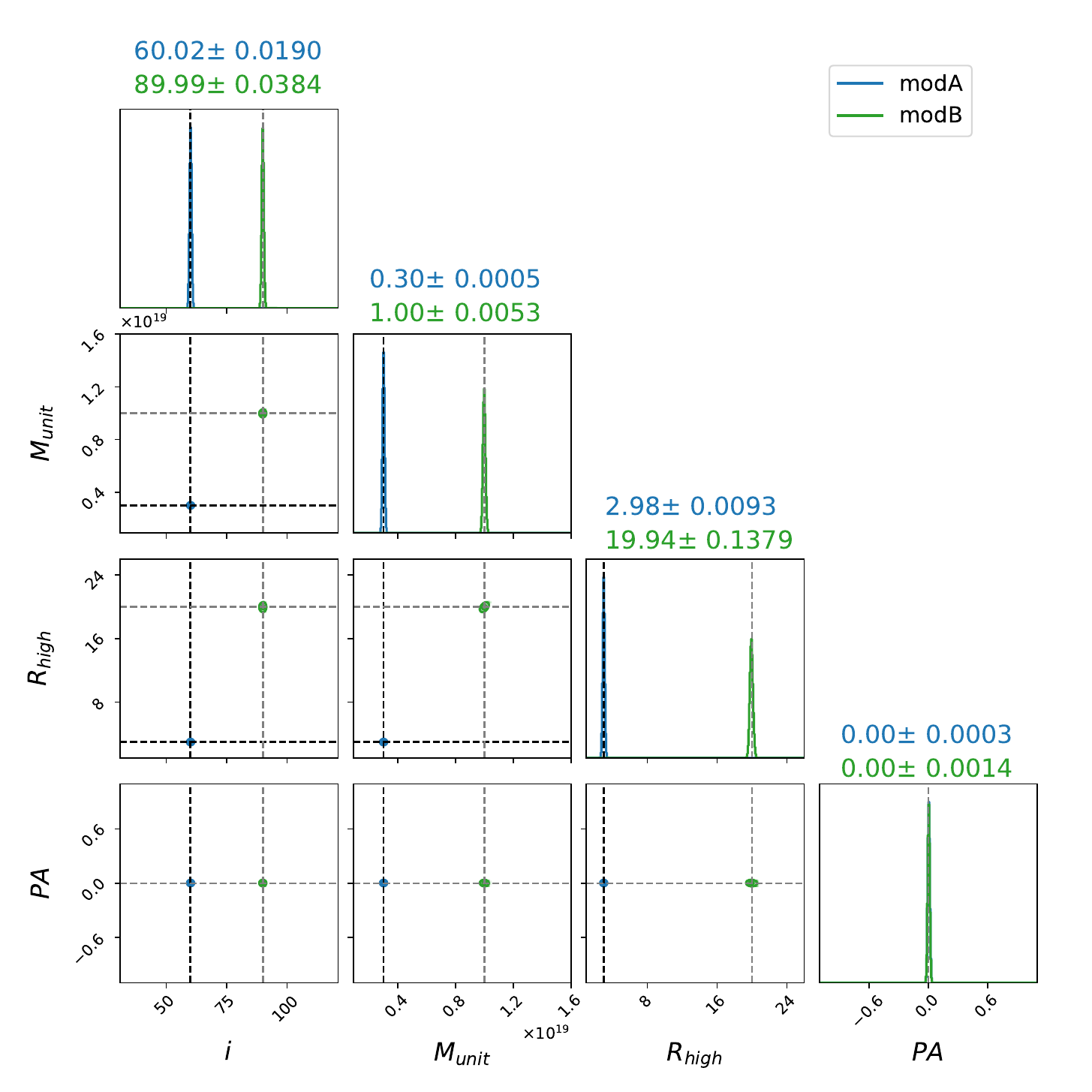}\\
  \caption{The triangle plot of four parameters estimation by fitting with the synthetic data based on models A and B. The main diagonal shows marginalized posterior distributions of all four parameters. The 6 plots in the lower left triangle show the joint densities for all the parameter combinations. The contours have been set to represent confidence of 0.68, and 0.9973 (1 and 3 $\sigma$). Dashed lines denote the truth (black for model A, gray for B). The lowest $\chi^2_{\rm eff}$ was 0.89 for both models.}
  %Middle: The log-likelihood evolution through out the run.
  %Bottom: The chains for the first 3500 MCMC of the simulation.
  %The burn-in for the PDs and the Gaussian fit values has been set to and 3500 MCMC. The total run time was 10000 MCMC.
  %The reported $\chi^2_{\rm eff}$ are the smallest values after the burn-in.} \ar{FIX THE BINS FOR ALL TRIANGLE PLOTS. WHEN POSTERIORS ARE NARROW USE PLENTY.} } 
   \label{fig:4P_methods}
\end{figure*}

Fig.~\ref{fig:4P_methods} shows a triangle plot of the posteriors and the joint probability densities of parameters given two different truths, models A and B. 
%The contour levels are for $0.68$ and $0.997$ confidence\monika{the info about contours is in the figure caption so does not need to be repeated here}. 
The burn-in window has been set to 3,500 MCMC steps. Our pipeline recovers both, significantly different, truth parameters. As visible in Table~\ref{tab:four-fit}, the numerical parameters of the four parameter fits together with 10,000 MCMC steps result in parameter estimation accuracy and precision comparable to those of the two parameter fits. 

\begin{table}
  \caption{Truth parameters and the pipeline performance for four parameter fitting with $1\%$ systematic error added to the simulated data, for a burn-in step of 3,500 MCMC. All parameters are sampled simultaneously, so there is only one $\chi_{\rm eff}^2$ value.}
  %\Shan{Is the table necessary?}
  %\RG{[Yeah point taken about redundancy, but I think it is nonetheless useful to show all inferred values next to their truth values in one place organized in a table.]}\ar{maybe now that xeff is not in plots, neither in text it makes more sense} }
    \centering
    \begin{tabular}{c|cccc}
         \hline 
         \hline
         %&\vspace{-0.8em}
         & truth & estimated &  $\chi_{\rm eff}^2$\\          \midrule
         \multicolumn{4}{c}{Model A}\\
         \hline %\vspace{-0.8em}\\
         PA($\pi$)  &0&$0\pm 0.0003$& $0.88$ \\
         %\hline
         $i(^{\circ})$  & 60 & 60.02$\pm0.02$&$0.88$\\
         $M_{\rm unit}(10^{18})$  &$3$&$3\pm0.0005$&$0.88$\\
         $R_{\rm high}$ &$3$&$2.98\pm0.01$&$0.88$ \\
         \hline %\vspace{-0.8em}
        \multicolumn{4}{c}{Model B}\\

         \hline
         PA($\pi$)  &0&$0\pm 0.0014$& $0.82$ \\
         %\hline
         $i(^{\circ})$  & 90 & 89.99 $\pm 0.0378$&$0.82$\\
         $M_{\rm unit}(10^{19})$  &$1$&$1\pm0.0052$&$0.82$\\
         $R_{\rm high}$ &$20$&$19.93\pm0.1345$&$0.82$ \\
         \midrule
    \end{tabular}
      \label{tab:four-fit}
\end{table}

\subsection{Effects of time variability on the parameter estimation}
\label{subsec:variability1}

\begin{figure*}
\begin{center}
 \includegraphics[width=0.921\linewidth,trim={0.1cm 2.0cm 5.0cm 2.2cm},clip]{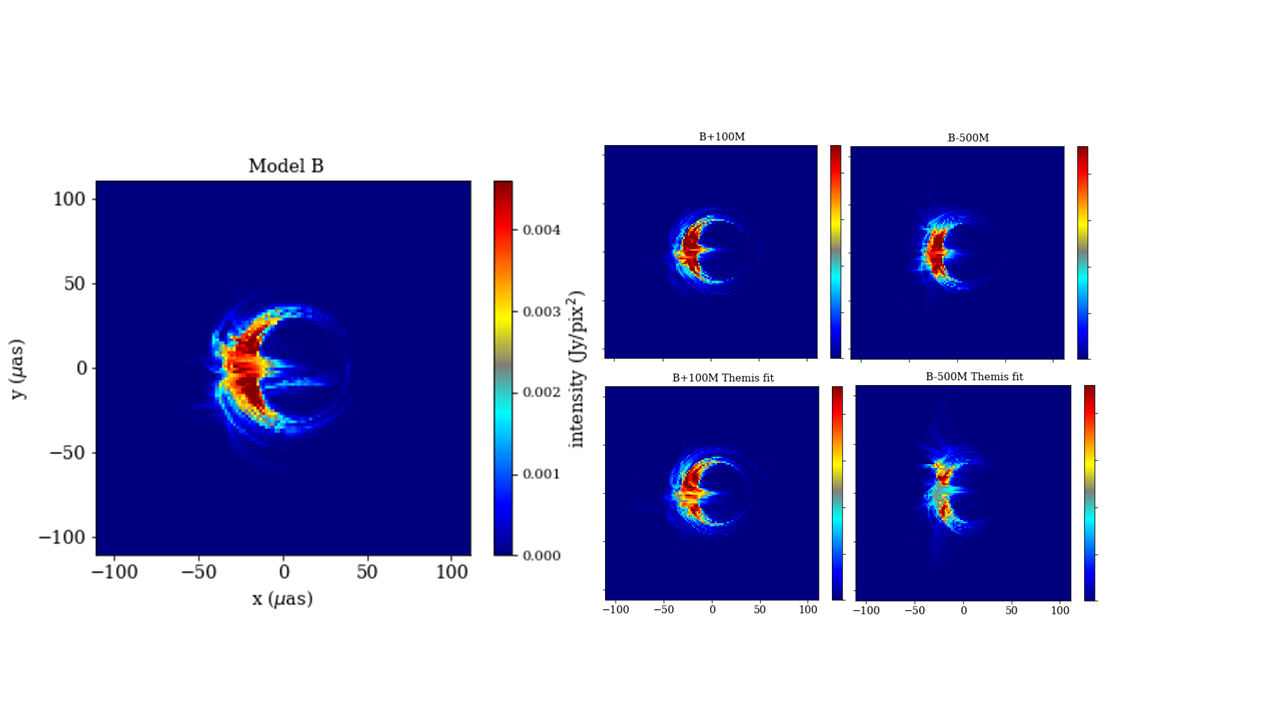}
  %\includegraphics[width=0.441\linewidth,trim={0.2cm 0.0cm 0 0.2cm},clip]{figures/ModB.png}
  %\includegraphics[width=0.221\linewidth,trim={0.2cm 0.0cm 0 0.2cm},clip]{figures/ModB+100.png}
  %\includegraphics[width=0.221\linewidth,trim={0.2cm 0.0cm 0 0.2cm},clip]{figures/ModB-500.png} 
  %\includegraphics[width=0.221\linewidth,trim={0.2cm 0.0cm 0 0.2cm},clip]{figures/ipole_moB+4k.png}%\hspace{0.2cm}
  %\includegraphics[width=0.221\linewidth,trim={0.2cm 0.0cm 0 0.2cm},clip]{figures/ipole_moB_avg.png}

  %\includegraphics[width=0.321\linewidth,trim={0.2cm 0.0cm 0 0.2cm},clip]{figures/ipole_moB+4k_themfit.png}
  %\hspace{5.6cm}  
        \caption{
    230\,GHz images of GRMHD model of Sgr~A* from different snapshots with the same (left and right top panels) and fitted (right bottom panels) GRRT parameters. The color codes the emission intensity (Stokes $\mathcal{I}$). A comparison between physical parameters and snapshots is presented in Table~\ref{tab:models_I}.}\label{fig:ModB_plot}
\end{center}
\end{figure*}

Here we examine the behavior of the pipeline when taking into account a possibility that the source may be changing in time (which is certainly true for Sgr~A* over a single night and for M87* over timescales of a week). In fact, our model image does not change with time yet (although this can be naturally incorporated in the future). Instead we examine the effects if the realization of variability in the model is different from the one in the data. We do so by fitting synthetic data from a model in a certain point in the simulation to synthetic EHT data created using a different time moment of the same simulation. model B+100M, has been created from the same GRMHD simulation and the same radiative transfer parameters, but 100 $GM/c^3$ later than model B, and similarly model B-500M, 500 $GM/c^3$ earlier than model B. Note that GRMHD snapshots are known to become sufficiently uncorrelated when separated by $20-30$M and so the adopted time offsets can be considered significant  \citep{Georgiev_2022}. The snapshots of the GRMHD model for the same radiative transfer parameters at different time moments are shown in Figure~\ref{fig:ModB_plot}. We still use model B for synthetic data, and B+100M or B-500M as a template for fitting. Our goal is to assess: (i) how poor the fit quality gets for a given error budget and (ii) how large the bias can be. 

\begin{figure*}
  \centering
  \includegraphics[scale=0.57,trim={0.1cm 0.5cm 0.5cm 0.1cm},clip]{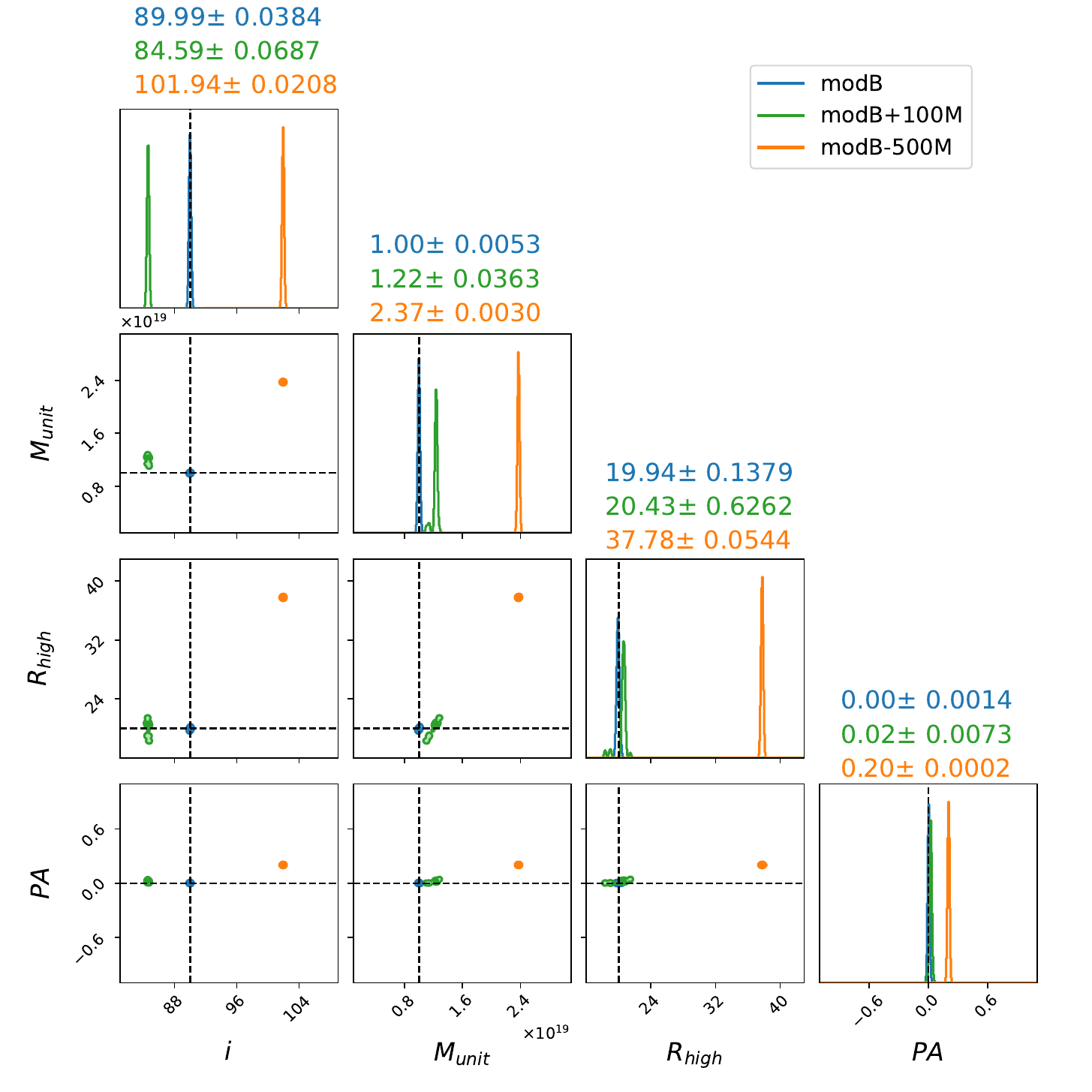}

  \caption{Same as Figure \ref{fig:4P_methods} for modB, modB+100M and modB-500M (all with $1\%$ systematic error). The corresponding $\chi^2_{\rm eff} = 0.89, 150, 181$ for the 3 models respectively.} 
   \label{fig:4P_modB}
\end{figure*}

Fig.~\ref{fig:4P_modB} shows parameter estimation for both snapshots.Regarding B+100M (green line), from the triangle plot it is clear that 2 parameters ($i, \,M_{\rm unit}$) have distributions with peaks shifted from the truth ($7\%$ and $23\%$, or 87$\sigma$ and 6$\sigma$ away from the truth, respectively), but in a coherent manner in a sense that the more distant snapshot is further away from the truth. Notice that in some cases the $3\sigma$ contours are on the edge of the truth (intersection of black lines). As for the two remaining parameters ($R_{\rm high}$ and PA), the distributions are able to cover the truth, perhaps a coincidence or an effect of time correlation given the poor fit quality (as expected in absence of inflated error budgets); in this test the $\chi^2_{\rm eff}=150.4$. 
% this is next section
%We will next consider an inflated error budget that captures the poor fit quality due to excess unaccounted-for variability.

In the same figure the orange line shows the 
parameter estimation for snapshot B-500M. It is evident that in this test the fit is unable to cover the truth in all four parameters, including PA. The log-likelihood is larger compared to that when fitting model B+100M, but $\chi^2_{\rm eff} =181$ stays roughly at the same level. 

The two tests above illustrate that on top of the expected poor fit quality a large bias is typically introduced into parameter estimation due to the intrinsic source variability. Table~\ref{tab:models_I} collects the best-fit parameters for the two runs. In addition, Fig.~\ref{fig:ModB_plot} shows the best-fit model images for the two tests above. The B+100M and B-500M best-fit images look somewhat different compared to model B (shown in the left panel) but overall crescent shape of the emission region is preserved. %\RG{[Do the differences remain when blurred to EHT beam? They will probably look more similar. We could compute MSE, DSSIM, NXCORR]}

\begin{table}
  \caption{The difference in GRMHD snapshots and physical parameters between the models presented in fitting test \ref{subsec:variability1}. The labels -100M, +500M, refer to how far a snapshot of the GRMHD simulation is with respect to model B (T=0). Themis fit, refers to the best parameters from the MCMC sampler. For B+100M and B-500M the sampler was fitting B with the aforementioned templates. The percentile in the fitted models note systematic error level. }
  
    \centering
    \begin{tabular}{c|ccccc}
         \hline\hline
         \vspace{-0.8em}\\\vspace{0.4em}
         Model & i & $M_{\rm unit}$ & $R_{\rm high}$& PA $[\pi]$& $\chi_{\rm eff}^2$ 
         \\\hline\vspace{-0.8em}\\
         %A  & 10680 M & $60$ &$3\times10^{18}$& $3$  \\
         %B  &10680 M &  $90$& $10^{19}$& $20$\\
         B+100M & $90$& $10^{19}$& $20$ & 0 & -\\
         B-500M & $90$& $10^{19}$& $20$ & 0 & - \\
         B+100M Themis fit $(1\%)$ & $84.5$&  $1.23\times10^{19}$& $20.5$ & 0.02& 150\\
         B-500M Themis fit ($1\%$)& $101.9$& $2.37\times10^{19}$& $37.8$ & 0.2& 181 \\
         B-500M Themis fit ($10\%$)& $90.0$& $1.0\times10^{19}$& $20.0$ & 0.0& 0.71 \\
         B-500M Themis fit ($30\%$)& $89.8$& $1.0\times10^{19}$& $20.3$ & 0.0& 0.41 \\
         \hline %\vspace{-0.8em}
         \hline
    \end{tabular}
      \label{tab:models_I}
\end{table}

\begin{figure*}
\begin{center}
    \includegraphics[width=0.921\linewidth,trim={0.1cm 0.1cm 0.1cm 0.1cm},clip]{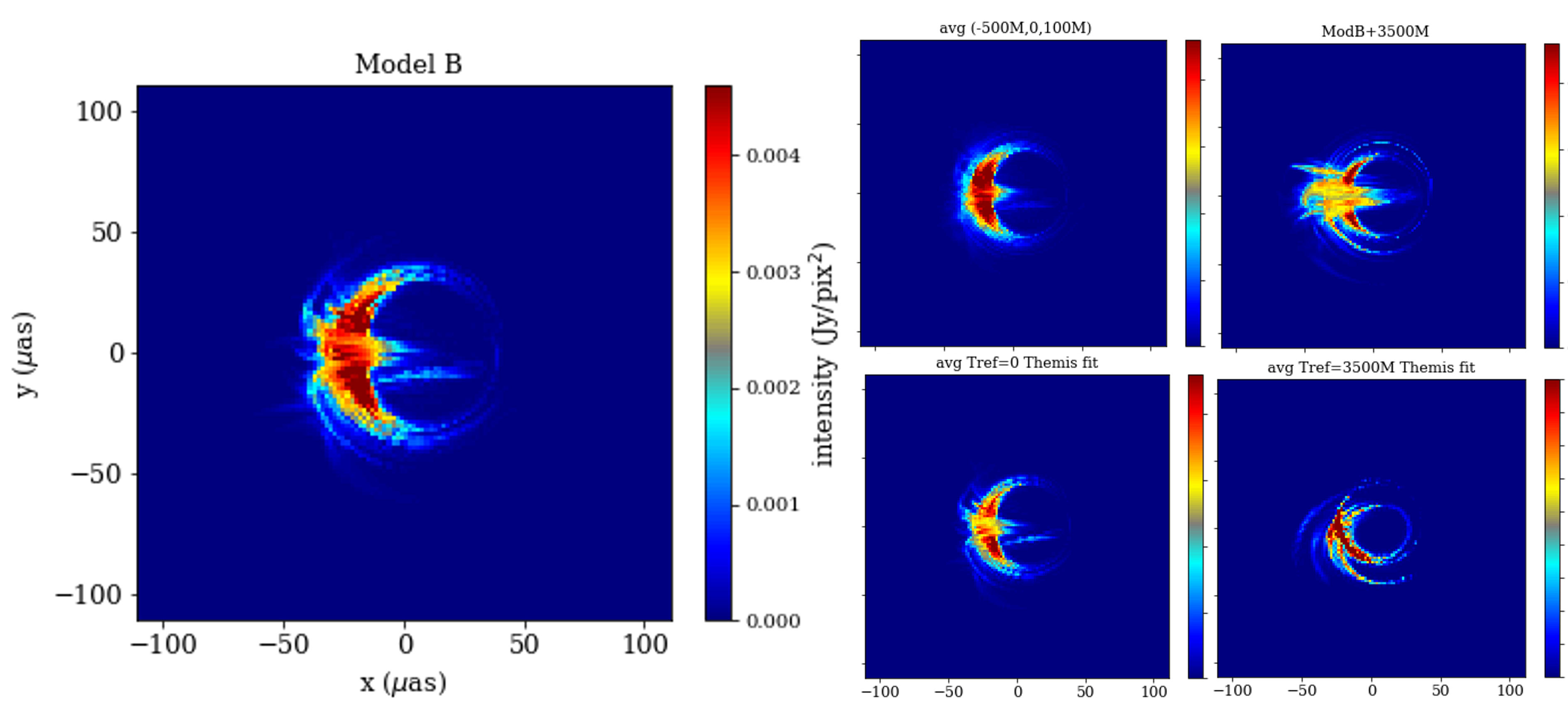}
  %\includegraphics[width=0.321\linewidth,trim={0.2cm 0.0cm 0 0.2cm},clip]{figures/ModB_Themfit+100_syser1.png}
  %\hspace{1cm}\includegraphics[width=0.321\linewidth,trim={0.2cm 0.0cm 0 0.2cm},clip]{figures/ModB_Themfit-500M.png}\\
  %\includegraphics[width=0.321\linewidth,trim={0.2cm 0.0cm 0 0.2cm},clip]{figures/ipole_moB_avg_themfit.png}
  %\hspace{1cm}
  %\includegraphics[width=0.321\linewidth,trim={0.2cm 0.0cm 0 0.2cm},clip]{figures/ipole_moB_avg_themfit_14k.png}    
        \caption{
    230\,GHz images of GRMHD model of Sgr~A* from different snapshots and using the GRRT parameters dictated by the fitting algorithm. For B+100M and B-500M the sampler was fitting B with the aforementioned templates. For avg models, Tref corresponds to the template used by the sampler. The color codes the emission intensity (Stokes $\mathcal{I}$). A direct comparison between physical parameters and snapshots can be seen in Table \ref{tab:models_avg}. Model avg Tref =3500M Themis fit introduces the strongest biases. The most striking ones to the eye are $M_{\text{unit}}$ and PA. The smaller $M_{\text{unit}}$ is a reaction to the extended nature of the snapshot, while the $PA=36\,$deg result of the different symmetry of the source. Note, that we fit CP which is particularly sensitive to symmetries on the sky.}
        \label{fig:ModB_themfit_plot}
        \end{center}
\end{figure*}

\subsection{Tackling variability: inflated errorbars and snapshot averaging}
\label{subsec:variability2}

In this section, we carry out two additional tests which may be useful when designing strategies on how to tackle the variability issues. 

%Lastly, we expand the variability investigation in two directions. First we explore a different error budget for the data of the truth. Second we create an averaged image using 3 snapshots from the simulation, and using this image as the truth we carry out snapshot fitting to synthetic data based on averaged image. These tests aim to probe an avenue to real data fitting, since we expect the real data to be smoothed out (like an average) and the first step to tackling the variability issue is simply inflating the errors of the data points to allow for more realisations of a particular snapshot. 

\begin{table*}
  \caption{The difference in GRMHD snapshots and physical parameters for fitting test \ref{subsec:variability2}. Model avg was made using snapshots -500M, 0 and +100M. Themis fit, refers to the best parameters from the MCMC sampler. Tref corresponds to the template used by the sampler. The percentile in the fitted models note systematic error level. }
    \centering
    \begin{tabular}{c|ccccc}
         \hline\hline
         \vspace{-0.8em}\\\vspace{0.4em}
         Model & i & $M_{\rm unit}$ & $R_{\rm high}$ & PA $[\pi]$ & $\chi_{\rm eff}^2$
         \\\hline\vspace{-0.8em}\\
         %A  & 10680 M & $60$ &$3\times10^{18}$& $3$  \\
         %B  &10680 M &  $90$& $10^{19}$& $20$\\
         B+3500M& $90$& $10^{19}$&$20$ & 0&- \\
         avg (-500M,0,100M) &  $90$& $10^{19}$& $20$&0&- \\
         avg (-500M,100M) Tref=0 Themis fit ($10\%$) & $96.04$& $0.98\times10^{19}$& $21.4$& 0.03&4.3 \\
         avg Tref=0 Themis fit ($10\%$) & $88.74$& $1.17\times10^{19}$& $26.9$& -0.06&10.2 \\
         avg Tref=3500M Themis fit ($10\%$)& $137.96$& $0.54\times10^{19}$& $40.0$&-0.6 &30.0\\
         \hline %\vspace{-0.8em}
         \hline
    \end{tabular}
      \label{tab:models_avg}
\end{table*}

%\RG{[The next sentence does not make sense as written. In fact, I'm trying to understand what the intended statement should be. A snapshot *is* a realisation of a model. I guess we mean that snapshots fit the data easier, because the larger error makes the analysis more permissive, but this needs to be clearer.]}
%Inflating the systematic errors can allow for more realisations of a certain snapshot, increasing the chance that the posteriors cover the truth. 

\begin{figure*}
  \centering
  \includegraphics[scale=0.57,trim={0.1cm 0.5cm 0.5cm 0.1cm},clip]{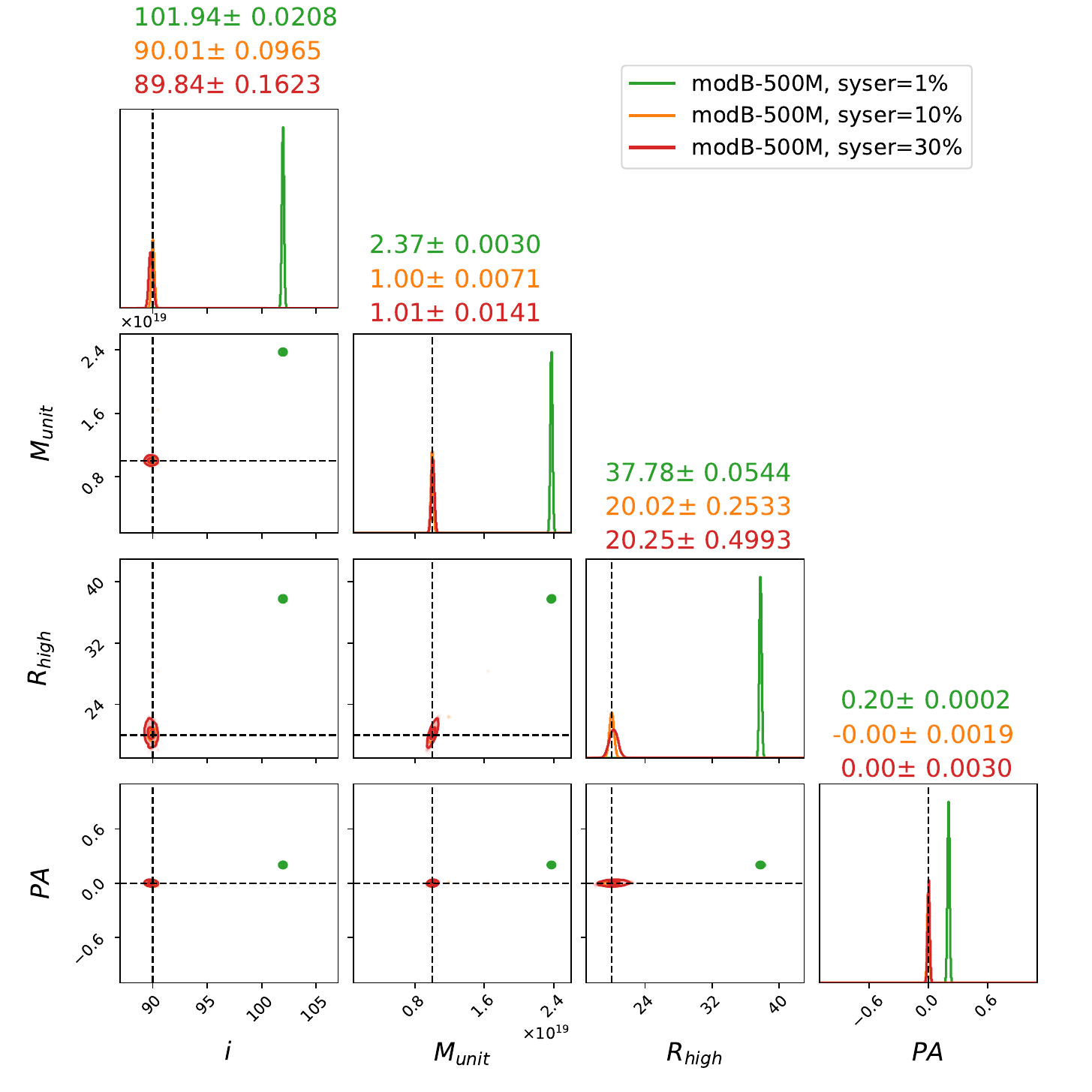}

  \caption{Same as Figure \ref{fig:4P_methods} for modB-500M ($1\%$ systematic error), modB-500M ($10\%$ systematic error) and modB-500M ($30\%$ systematic error). The corresponding $\chi^2_{\rm eff} = 179.5, 0.71, 0.41$ for increasing error budgets ($1\%$, $10\%$, $30\%$). The snapshots from all models (plus model B) are visible bellow, in Fig. \ref{fig:modb_se_all}.}
   \label{fig:4P_syser10}
\end{figure*}

The first obvious step to address variability impact on parameter estimation is to simply inflate the systematic errors of the data points to enable the analysis to be more permissive and fitting to be easier. Fig.~\ref{fig:4P_syser10} demonstrates model B fitting using snapshot B-500M 
with 3 different choices of systematic error, namely: $1\%$, $10\%$ and $30\%$. Already with systematic errors of $10\%$, the fitting of snapshots separated by $\Delta t=$500M recovers the truth. This test validates that the pipeline can find the truth even with snapshot misspecification, a necessary feature for real data fitting, where most certainly all our models will be (at best) only approximations to the real image. At $30\%$ errors, the posterior distributions of all free parameters widen (as expected), notice that $\chi^2_{\rm eff} = 179.5, 0.71, 0.41$ for increasing error budgets, while still covering the truths at $1\sigma$ confidence. 
%This analysis if further supported by the corresponding $\chi^2_{\rm eff} = 179.5, 0.71, 0.41$ for increasing error budgets ($1\%$, $10\%$, $30\%$). 
For this specific example, the $30\%$ and arguably $10\%$ cases are slightly overfitted, presumably due to an overestimated systematic error to capture the intrinsic variability. Note that for general and realistic cases this may be slightly different, but this needs to be investigated more thoroughly in a dedicated future work.

\begin{figure*}
\begin{center}

  \includegraphics[width=0.932\linewidth,trim={0.2cm 6.0cm 0 4.6cm},clip]{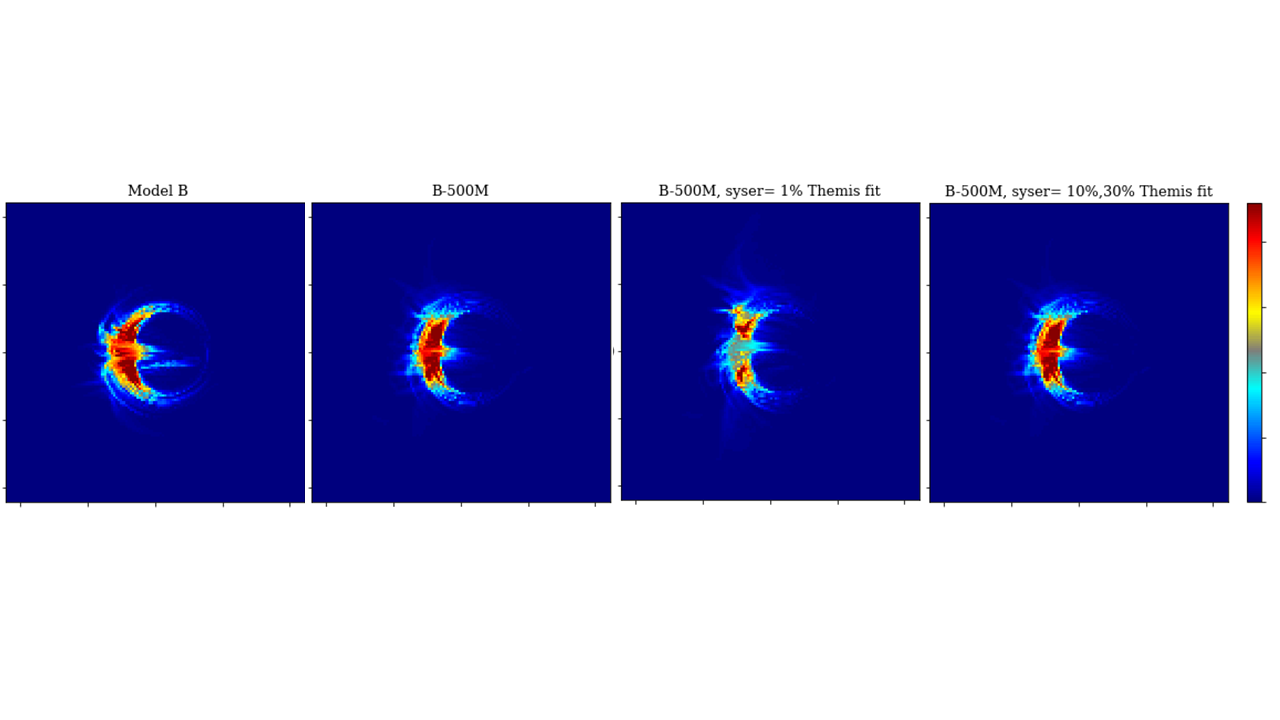}
        \caption{
    230\,GHz images of GRMHD models from Fig. \ref{fig:4P_syser10}. The figure aims to underline the improvement on the last two panels, going from $1\%$ errors, to $10\%$ and $30\%$. }
        \label{fig:modb_se_all}
        \end{center}
\end{figure*}

Sgr~A* is changing on timescales that are short compared to a full night of EHT observation. To be precise $500$M for Sgr~A* is equivalent to $165$min.
In this case VLBI data collected over one full night represent a smoothed-out image of a varying accretion flow.
%Additionally, we test our fitting scheme under the hardest challenge for real data, which is the fact that VLBI data represent a smoothed-out image of the accretion flow around a super massive black hole, in particular for the case of Sgr~A*, where the typical timescales are much shorter than a full night of observation. 
%To be precise $500$M for SgrA* correspond to $165$min, meaning the EHT array is scanning over different snapshots for a single image. 
To emulate such smoothing effect in our pipeline we create the truth synthetic data by averaging 3 snapshots (models: B-500M, B and B+100M), called model avg. We then fit a single snapshot to the "averaged" truth (shown in Figure~\ref{fig:ModB_themfit_plot}, left panel). We consider two cases: first where the fitted snapshot is a part of the averaged image (model B, shown in Figure~\ref{fig:ModB_themfit_plot}) and second where the fitted snapshot is approximately 3,500M away from the average image (Figure~\ref{fig:ModB_themfit_plot}). These two fits are called "avg Tref=0" and "avg Tref = 3500M" respectively. As a sanity check we also perform a fit with model avg (-500M,100M), that excludes the snapshot Tref=0 from the truth. The systematic error added to the synthetic data was $10\%$ for all cases.

\begin{figure*}
  \centering
  \includegraphics[scale=0.57,trim={0.1cm 0.5cm 0.5cm 0.1cm},clip]{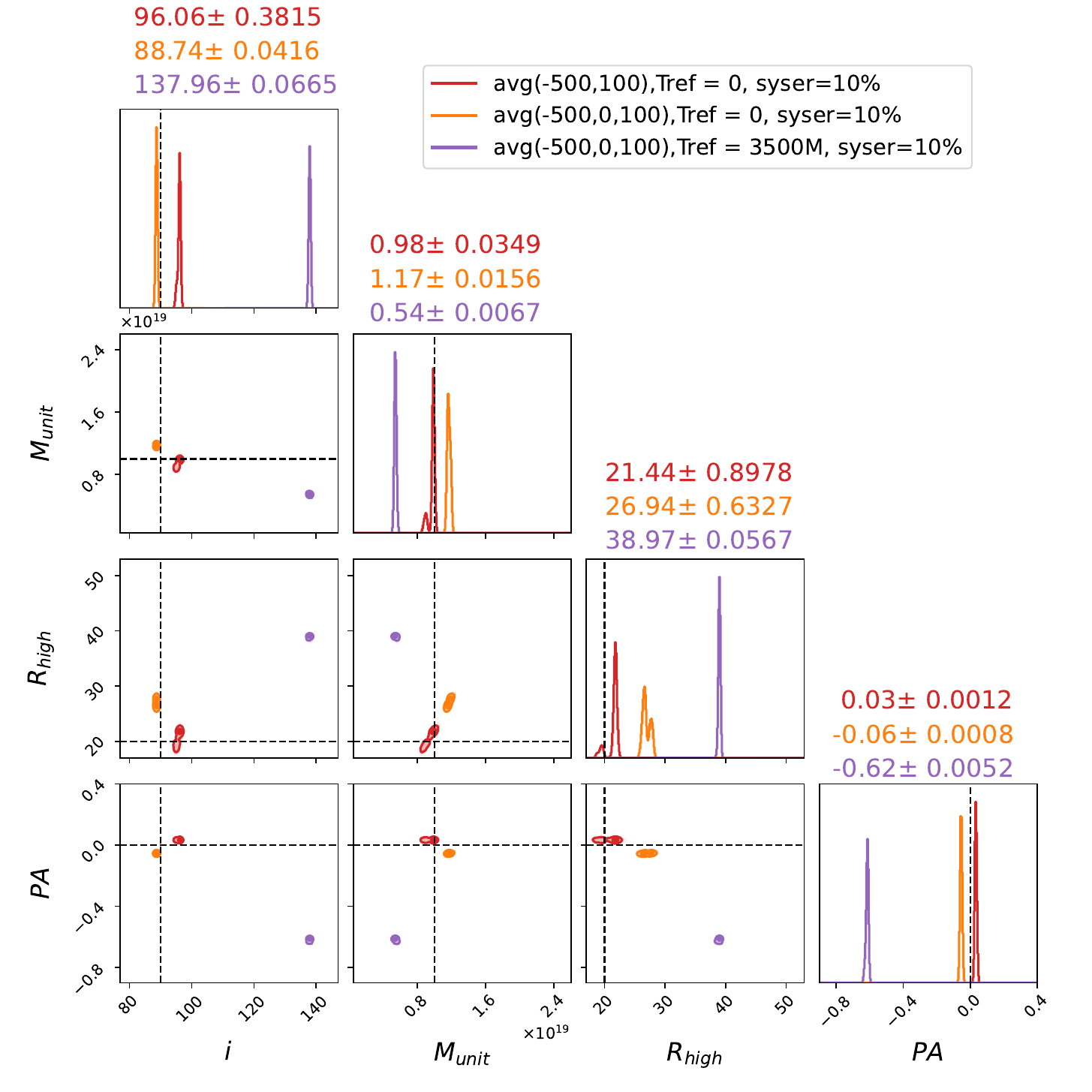}
  \caption{Same as Figure \ref{fig:4P_methods} for models avg(-500,100), Tref=0, avg(-500,0,100), Tref=0 and avg(-500,0,100), Tref=+3500M, all with $10\%$ systematic error. The corresponding $\chi^2_{\rm eff} = 4.3, 10.2, 30.0$ for the three models respectively.}
   \label{fig:4P_avg}
\end{figure*}

The results of this fitting exercise are shown in Figure~\ref{fig:4P_avg} and their parameters are listed in Table~\ref{tab:models_avg}. The model avg Tref=0 (orange line) converged to values close to the truth with the largest deviation for $R_{\text{high}}=26.9$ and total $\chi_{\rm eff}^2=10.2$. The produced image avg Tref=0 Themis fit is visible in Figure~\ref{fig:ModB_themfit_plot} (bottom right panel) closely resembles the averaged image (the cross-correlation between them is $0.972$ using the NXCORR tool within \ehtim software). The fit from model avg (-500M,100M), Tref=0, is visible in Fig. \ref{fig:4P_avg} (red colour). The results are of matching quality with similar biases and $\chi_{\rm eff}^2=4.3$, ensuring that the smoothing of the truth is more significant than the existence of Tref=0 in the averaged truth image.

In the second case of model avg Tref=+3500M the biases are significantly larger, placing an upper limit in snapshot misspecification at roughly 500M. The fitted image (avg Tref=3500M Themis fit) is visibly different from the averaged image, but explains some of the posterior values, such as the low $M_{\text{unit}}$ to make emission narrower, and negative PA to roughly match the emission region asymmetry on the sky. 

To sum up, both these tests suggest that variability will  play a detrimental role in parameter estimation when moving to real data fitting. However, a more sophisticated implementation of the noise models arising from variability studies \citep{Broderick22_variability,Georgiev_2022}, both on baseline and time domain, could prove extremely fruitful, and we plan to examine this in a future study.

\section{Discussion and Conclusions}
\label{sec:discussion}

% this section should contain
% \begin{itemize}
% \item short summary of the paper
% \item shortcomings of the present approach described in a positive way
% \item possible extensions: e.g. time variability, polarization
% \item Successes of the current work and future plans? 

We created a pipeline toward Bayesian inference by fitting the GRMHD models to EHT observations and estimate model parameters. Similar efforts have been previously made by \citet{2016ApJ...832..156K} where sampling of GRMHD images was done only using two parameters: PA and total flux normalization. Also in \citet{2022ApJ_march} a new MCMC algorithm was introduced for sampling of geometric, crescent models for image features (such as the shadow radius, the width of the ring etc.). In \cite{2023ApJ_primo} a PCA-based image reconstruction was developed, using an ensemble of simulated GRMHD images for fitting VLBI data. Lastly, in \citet{Ale24} image moments were used to characterize GRMHD snapshots as a means for model discrimination. Here we sample multiple parameters which require radiative transfer calculations in every MCMC-step which is a significant leap compared to previous work. We tested the pipeline over two distinct models (A,B) with differing inclination angle and $R_{high}$, but more importantly, we made first steps towards tackling the time variability issue of such systems. The main results of the current work lies in Fig.~\ref{fig:4P_modB}, where we showed that with the correct consideration of error budgets the pipeline is capable of retrieving correct parameters even for misspecified snapshots. In Fig.~\ref{fig:4P_avg} we showed that the miss-specification can work even for an averaged truth from multiple snapshots. Of course, that does not come without limitations as for the same averaged snapshot with a fitting template 3500M away, the pipeline misses significantly the truth in all parameters. Template spacing of 1000M, or 500M to be more moderate, could potentially solve that and it would decrease the necessary snapshots by an order of 100, from 500 (cadence 10M) used in AIS to 5-10 (cadence 1000M-500M) with our scheme.      

In this stage we focused on fitting models to observed closure phases constructed from interferometric visibility phases. If one chooses to also fit visibility amplitudes, another thing that should be taken into account is that in case of Sgr~A* (but not M87*), the visibility function should be additionally modified to include smearing effects caused by refractive scattering of radio waves by free electrons in the galaxy \citep{2015ApJ...805..180J}, causing artificial small-scale substructure in the image.

Despite being an improvement on static libraries, the adaptive parameter estimation is still time-consuming. In particular, to run the multi-parameter fit on a university cluster, with 60 CPU cores, 1 GB per core, for 10,000 MCMC, takes approximately 170 wall-clock hours. In the same cluster running the snapshot scoring (part of AIS) takes 20 wall-clock hours with 24 cores, for a certain parameter combination (so using only 5 inclination values and 5 $R_{\rm high}$ values takes 500 hours). For a usage of $\sim10$ snapshots our method is already faster, plus the added value of not having to create, save and manage the millions of snapshots required for AIS. 

Another time-consuming part (for both the static library scoring and our approach) is the GRMHD simulation itself. At present we do not consider different GRMHD simulations and only call the ray-tracing code to generate different model images from the same simulation. The parameters which we are interested in the GRMHD, such as the spin of the black hole, could not be estimated under the current settings. Applying the current method to multiple simulations is a first, direct way forward leaving a model selection problem that could be tackled with Bayesian evidence or information criteria. How to simplify the model and generate the model image faster is a big challenge and that is the reason why fitting simple phenomenological models to observations is another practical way to compromise at present, such as \citet{1995ApJ_adaf} and  \citet{2006_riaf} or the more modern approaches of \citet{2022_kerrbam}, \citet{2024_jukebox} and \citet{2024_bipole}. 

The MCMC sampling part is fast due to the highly parallel development. The computing performance could be improved by carefully choosing the numerical parameters (e.g. the temperature, the number of walkers) to be better adapted to the computer. Another bottleneck arises from load imbalance on the radiative transfer side. At each tempering level independent model images are being generated for different parameter values, some of which such as higher $\dot{M}$ (ie. higher density and opacity) will take longer to compute than others. On this front, recent developments in ray-tracing optimization, such as the GPU version of \ipole presented in \cite{MoMo_GPU_2023} can be proven useful for the speed-up of the pipeline. 

We have presented and validated the first Bayesian scheme to infer properties from GRMHD simulations from their simulated model images and visibility data as measured by an EHT-like VLBI configuration. This is a major step in fully utilizing the predictive power from GRMHD simulations of accreting black holes which previously have only been compared to VLBI data in more indirect ways and by using a-priori fixed parameter surveys. The work presented here eliminates simplifying scaling assumptions with total flux and BH mass in previous EHT VLBI analysis using GRMHD models \cite{eht2019e}. It further allows improved conclusions from GRMHD models given a VLBI dataset: (i) a refined inference in a continuous posterior distribution instead of discrete apriori chosen values (ii) efficient extension of the probed prior range (for instance beyond $R_{high}=160$) of the explored model parameters, which would otherwise become prohibitive with current strategies and (iii) a more efficient pathway to expand the parameter space to include any additional parameter that \ipole can vary. More advanced samplers can easily handle much higher-dimensional likelihood surfaces than will ever be explored with such models. Instead, the key improvement will be to speed up the evaluation of a single likelihood for instance by speeding up disk I/O. 
%\Shan{ And in the end of this paragraph, discuss about the benefits of expending the estimated parameter space, will improve the theoretical interpretation of the data. }

Much work is still needed to get the most out of such inferences, but the next steps (sampler improvements, better likelihood approaches using complex visibilities, GRMHD-informed error budget etc \cite{eht2022d} \cite{Broderick22_variability,Georgiev_2022} are both clear and already implemented in other analyses in \THEMIS. 
Furthermore, following the results of \cite{eht2021polar}, where polarization is resolved from M87, so called closure traces\footnote{Closure traces, introduced by \citealt{Broderick_2020}, is a data product constructed from polarimetric VLBI visibilities, that are immune to both station gains and polarization leakage (encoded in the so-called "D-term").} can be included in the pipeline, making the fitting of the polarization to full visibilities possible.
We envision that improved analysis schemes will greatly benefit the GRMHD community and theoretical interpretation of accreting black holes in the near future.

\section*{Acknowledgements}
We thank Paul Tiede, George Wong, Lia Medeiros and Michi Baub{\"o}ck for fruitful discussions. This publication is part of the project the Dutch Black Hole Consortium (with project number NWA 1292.19.202) of the research programme the National Science Agenda which is financed by the Dutch Research Council (NWO).
{\it Software used in the paper:} \ipole, \THEMIS, \ehtim, Python.

\bibliographystyle{mnras}
\bibliography{sample} % if your bibtex file is called example.bib

%%%%%%%%%%%%%%%%%%%%%%%%%%%%%%%%%%%%%%%%%%%%%%%%%%

%%%%%%%%%%%%%%%%% APPENDICES %%%%%%%%%%%%%%%%%%%%%

\appendix

\section{Optimal image resolution}\label{app:image_res}

%add 256 and discuss why choose 128
% (a1) and (c1) should have y ticks
\begin{figure*}
  \centering
   \includegraphics[scale=0.387,trim={0.1cm 0.1cm 0.1cm 0.1cm}]{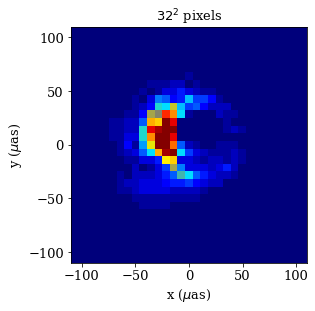}
   \includegraphics[scale=0.387,trim={0.1cm 0.1cm 0.1cm 0.1cm}]{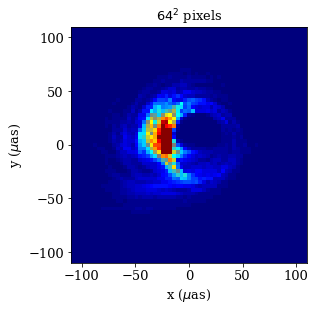}
   \includegraphics[scale=0.387,trim={0.1cm 0.1cm 0.1cm 0.1cm}]{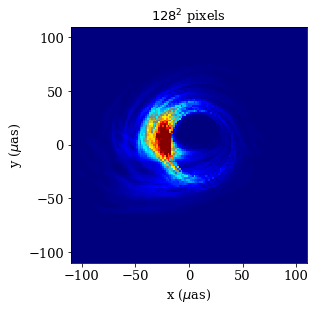}
   \includegraphics[scale=0.387,trim={0.1cm 0.1cm 0.1cm 0.1cm}]{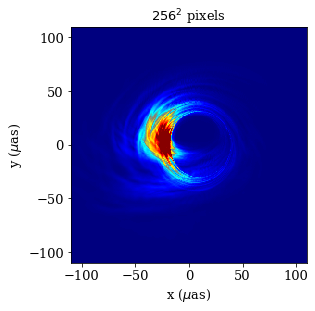}
   \includegraphics[scale=0.7,trim={0.1cm 0.1cm 0.1cm 0.1cm},clip]{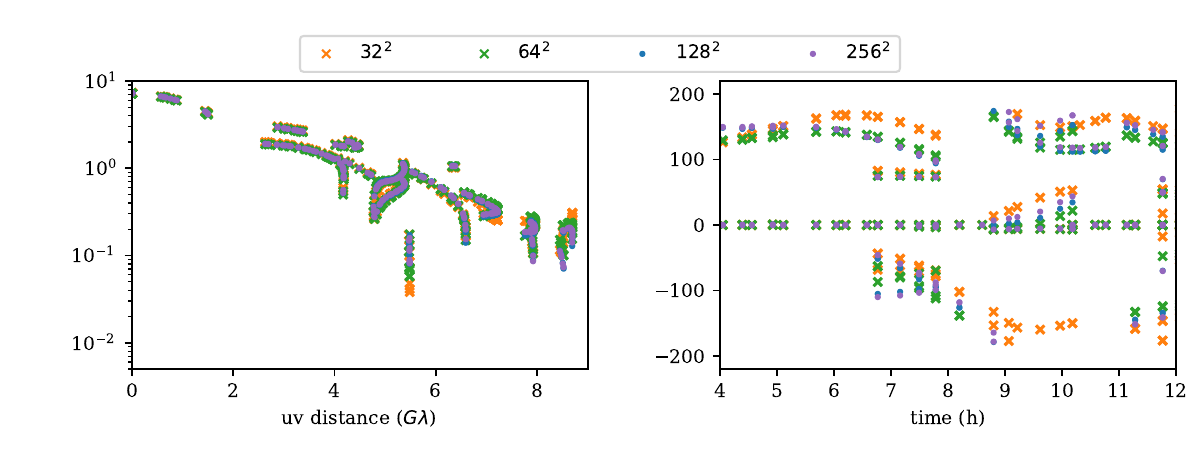}
  \caption{Top: the modeled images for 32$\times$32 pixels, 64$\times$64 pixels, 128$\times$128 pixels, 256$\times$256 pixels. Bottom: The left plot shows the visibility amplitudes for all different resolutions. The right one all the closure phases. The colors have been set to represent: blue ($32^2$), green ($64^2$), orange ($128^2$) and purple ($256^2$).   
  }
  \label{fig:entim_Themis}
\end{figure*}

\begin{figure*}
  \centering
  \includegraphics[scale=0.57,trim={0.1cm 0.5cm 0.5cm 0.1cm},clip]{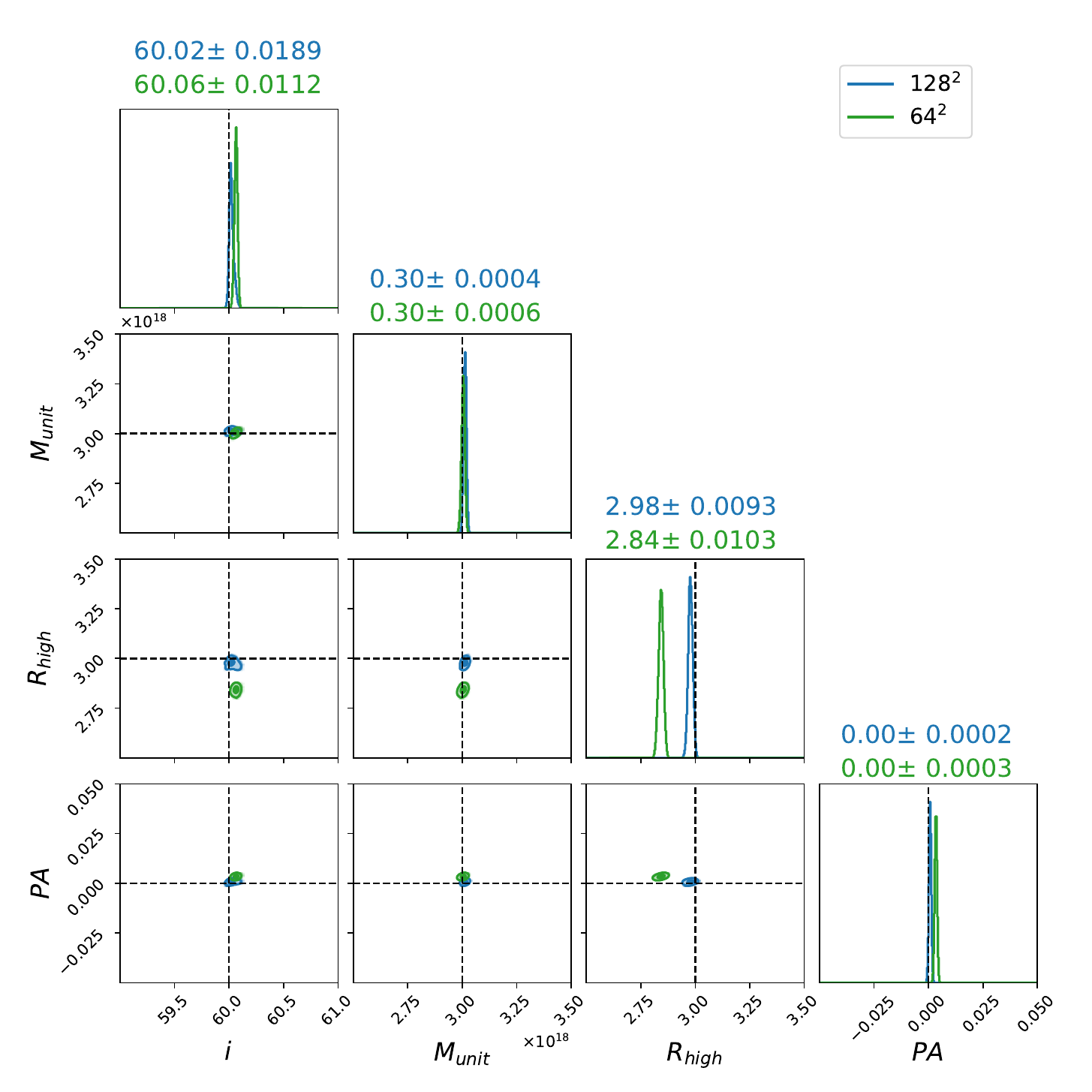}
  \includegraphics[scale=0.47]{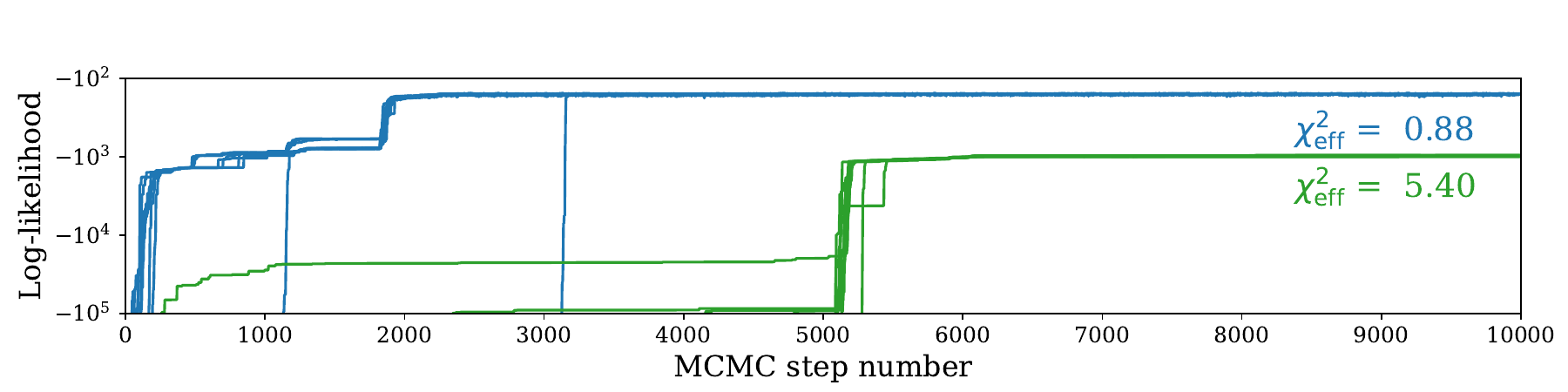}\\
        \includegraphics[scale=0.47]{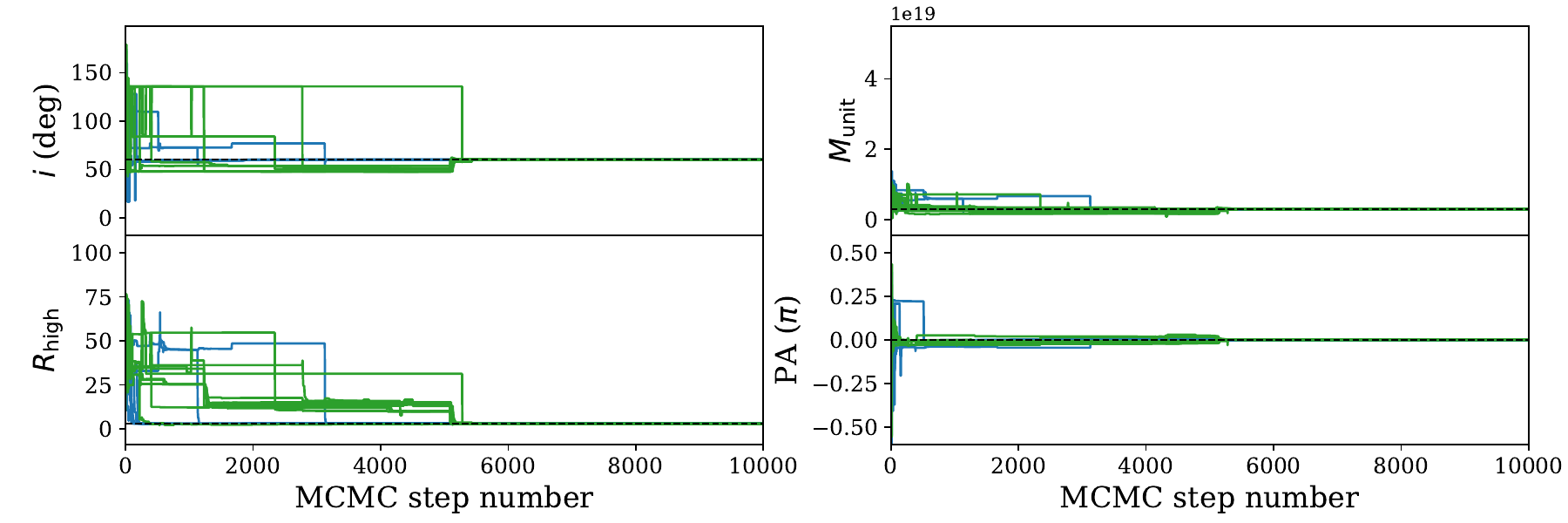}

  \caption{Top: Same as Figure \ref{fig:4P_methods} for model A using different resolution ($64^2$, $128^2$). 
  Middle: The log-likelihood evolution through out the run.
  Bottom: The chains for the whole duration of the MCMC run.
  The burn-in for the PDs and the Gaussian fits has been set to 6000 and the total run time was 10000 MCMC. The reported $\chi^2_{\rm eff}$ are the smallest values after the burn-in. In short: low resolution introduces poorer fit quality, but the posteriors are largely un-biased with the exception of $R_{high}$. The latter bias in $R_{high}$ is subtle in magnitude compared to the coarse spacing in the EHT libraries, but should be further investigated in subsequent work.}
   \label{fig:4P_64-128}
\end{figure*}

To reduce the computational cost of the pipeline we look for a minimum model image resolution for which the synthetic VLBI data are converged. 
Given the fiducial image of model A (with parameters $i=60^{\circ}$, $M_{\rm unit}=3\times 10^{18}$, $R_{\rm high}=3$, $R_{\rm low}=3$, PA=0), shown in Fig.~\ref{fig:synthetic_data}, we use \ehtim to produce the visibility amplitudes and closure phases for four different image resolution cases: 32$\times$32 pixels, 64$\times$64 pixels, 128$\times$128 and 256$\times$256 pixels. 
 
We compare all the observables generated by \ehtim, the results of which are plotted in Fig.~\ref{fig:entim_Themis}. 
%Eventually, what we want to prove is that even though the data get better with higher resolutions, the step from 128 to 256 is very small, and definitely is not worth the extra computational time, at this point at least.  
In the top row we can see the natural improvement of images with resolution. In the bottom one we show the VA and CP data for all resolutions. What we want to convey is that even though the data get better with higher resolutions, the step from 128 to 256 is small, while the steps before that large. 
Indeed, we see that points of green and orange show sizable divergence, but the blue points are often `hidden' in the plots, suggesting that they match very well with the points from $256^2$ (purple). 

To further strengthen our thesis and demonstrate the benefit of using data with resolution  $128^2$, in Fig. \ref{fig:4P_64-128} we compare the results from fitting model A, for all four parameters, using two different resolutions ($64^2$, $128^2$). The higher resolution outperforms the lower one in a number of fronts. The posteriors for $64^2$ are not covering the truth for all parameters ($R_{\rm high}$ is shifted). This is reflected in the likelihood graph as well, where in lower resolution the quality of the fit is not as good ($\chi^2_{\rm eff}=5.37$). Lastly, the lower resolution exhibits a slower convergence, that can be seen both in the chains and the likelihood development.  
%In the third panel we can see some points that reach differences of around 10 deg which is not insignificant. After investigation we have associated these data with a few triangles containing very long baselines, like AZ-SP-AA (Arizona, South Pole, ALMA) that exhibit a more sensitive action in resolution changes. 

Overall, it seems that a resolution of $128^2$ is necessary for sufficient fit quality while  resolution of $256^2$ does not provide further improvement at higher computational expenses.
\section{Dependence on choice of samplers, walkers, temperatures and initialization}\label{app:samplers}

\begin{figure*}
    \centering
    \includegraphics[width=0.715\linewidth,trim={0.7cm 0.0cm 0 0.2cm},clip]{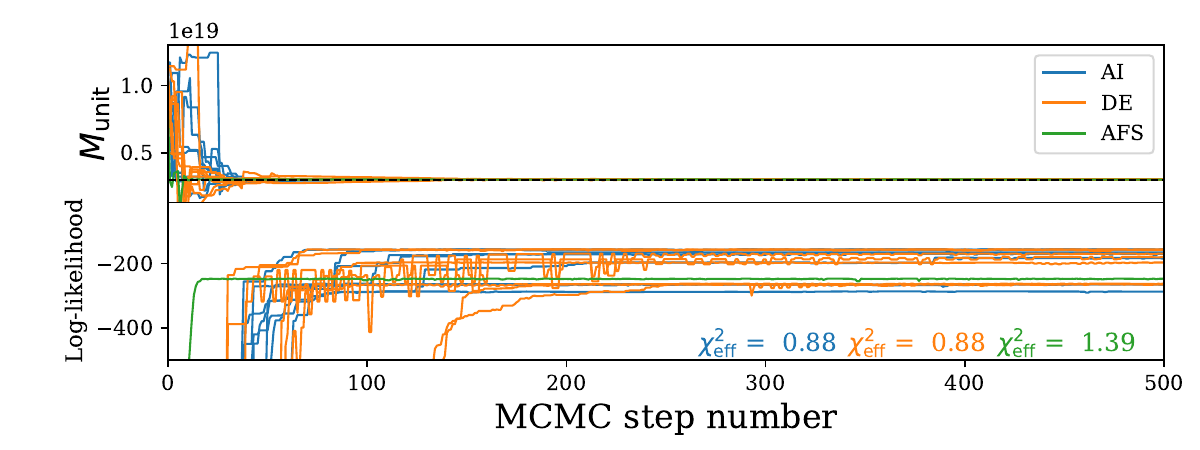}\\
    \centering
    \includegraphics[width=0.715\linewidth,trim={0.7cm 0.0cm 0 0.2cm}]{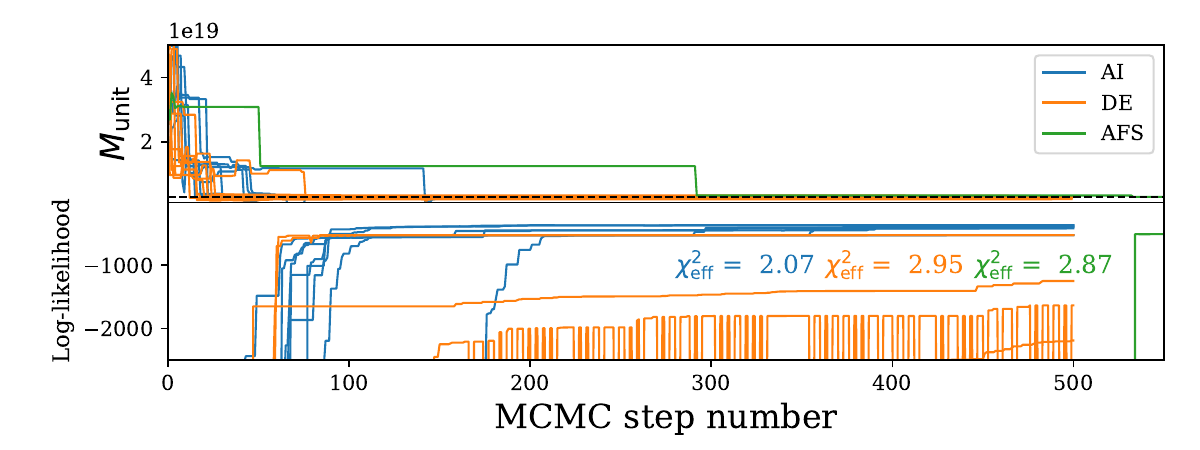}
    \caption{Trace of the chains fitting ${M_{\rm unit}}$ for different samplers (AI: affine invariant, DE: differential evolution, AFS: automated factor slice). The run shown in the top panel was initialized at $8 \times10^{18}$, while the bottom at $3\times 10^{19}$. Note that the last MCMC step in the bottom is 550. AI and DE were run with 40 cores and their computational efficiency was 1.5 steps per core per hour. AFS was run with 10 cores and the efficiency was higher at 2.5 steps per core per hour.} 
    \label{fig:Mu_samps}
\end{figure*}

\begin{figure*}
    \centering
    \includegraphics[width=0.715\linewidth,trim={0.7cm 0.0cm 0 0.2cm},clip]{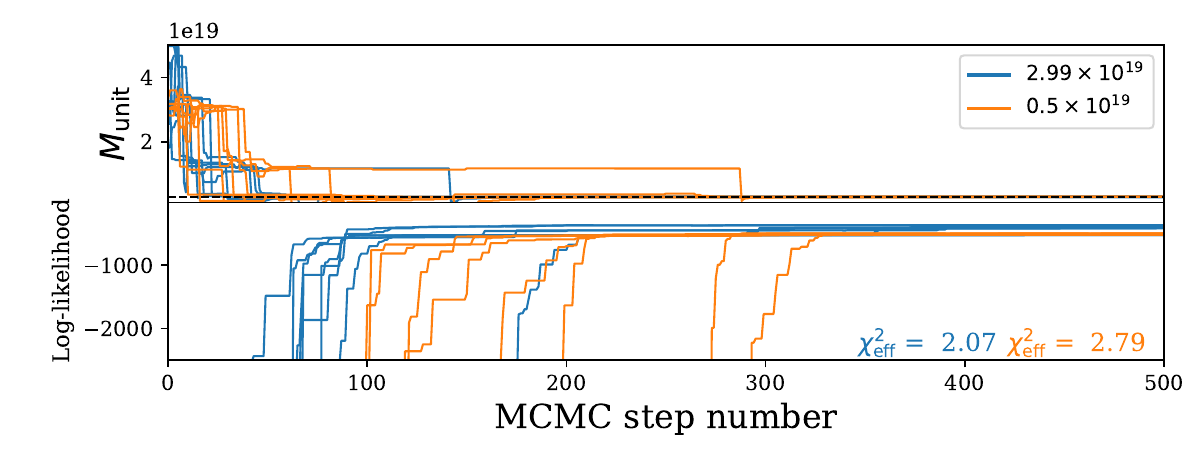}
    \centering
    \caption{Trace of the chains fitting ${M_{\rm unit}}$ for different initial ranges using the AI sampler. The initial value is $3 \times10^{19}$.}
    \label{fig:Mu_ranges}
\end{figure*}

In this appendix, we search for the optimal sampling method and chains initialization. We do this by various fitting procedures and direct comparisons for model A.

Within the parallel tempering sampler, an algorithm should be chosen to decide on the next step parameters. There is a handful of options implemented in \THEMIS, from which we consider three for this project. The  first one is the affine invariant (AI) method described in detail in \citet{Goodman2010CAMCoS5_65}. Under the affine transformation the ill-shaped density probability will not bring any extra difficulties as the simple single-variable MCMC samplers (e.g. Gibbs sampler) do. This method is widely used and well tested in the famous MCMC python package EMCEE \citep{Foreman-Mackey2013PASP125.306}. The second possible algorithm is the differential evolution (DE) method introduced by \citet{TerBraak2006S&C16.239} and developed further in \citealt{Nelson2014ApJS210.11}. Similarly to the affine invariant method each chain draws new values by using positions from other chains in the parameter space. The difference is that DE uses all chains per draw while AI only one. This can lead to discrepancies in their performance, depending on the problem. Due to the pipeline constructions, the number of processors should be at least $N_{\rm T}$ and be integer times of $N_{\rm T}\times N_{\rm W}/2$ for both samplers.  The last option is the automated factor slice (AFS) algorithm \citep{doi:10.1080/10618600.2013.791193}. It was implemented more recently than the others within \THEMIS and it has a key difference since it operates only with one chain, multiplied by the temperatures. It improves from the simple Metropolis-Hastings (MH) sampler by using an in-between step of redefining the sampling pool on every step.

Fig. \ref{fig:Mu_samps} shows the chains and the corresponding likelihood for all three samplers fitting 2 parameters ($M_{\rm unit}$ and PA). The top has been made with an initial value of $8\times10^{18}$ while the bottom with $3\times10^{19}$ (truth is at $3\times10^{18}$). In the top panel all samplers behave similarly, they converge towards the truth fast (with AFS being the fastest $\sim 10$ steps) and with similar $\chi^2_{\rm eff}$ values (AFS has a higher value that persists till after 5000 MCMC steps). Already we can notice that DE in contrast to AI is more jittery, probably to the fact that the number of chains and temperatures is too high for the level of communication DE sampler imposes.   

In the bottom panel the picture is different. AFS which was the faster sampler in the previous case, now converges to the truth last (at 530 MCMC steps). Further more, even though it seems like DE is faster than AI, that is actually not the case, since almost all chains are close to the truth but not exactly and vary significantly for a long period. All this is reflected in the likelihood plot, where the oscillation of DE chains is more evident.

As a last remark the computational efficiency of AI and DE is 60 and 70 steps per hour respectively, using 40 cores (8 chains and 10 temperatures, divided by 2), so $\sim1.5$ steps per core per hour while AFS is 25 steps per hour, using 10 cores (1 chain and 10 temperatures), so $2.5$ steps per core per hour. 
%Since the number of cores is not a problem at this level, the performance of AI and DE is to be preferred. \RG{[What do you mean by "It is not a problem"? You mean anyone has 40 cores available?]}

\begin{figure*}
    \centering
    \includegraphics[width=0.79\linewidth,trim={0.3cm 0.0cm 0 0.2cm},clip]{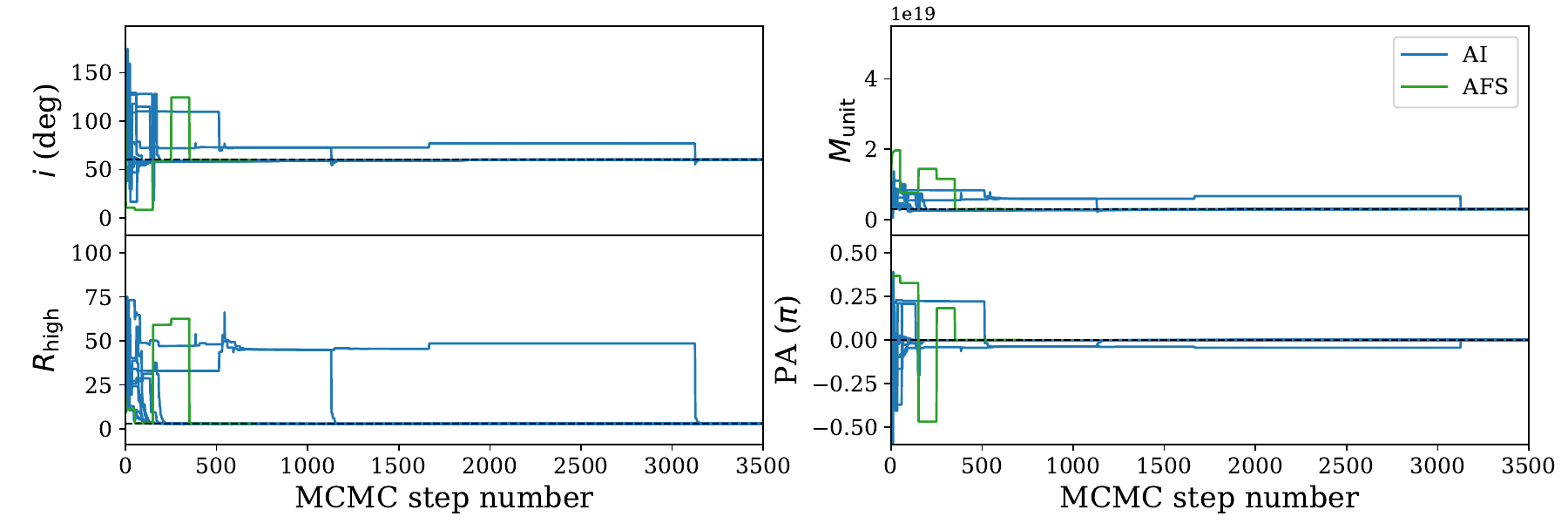}\\
    \centering
    \includegraphics[width=0.79\linewidth,trim={0.3cm 0.0cm 0 0.2cm}]{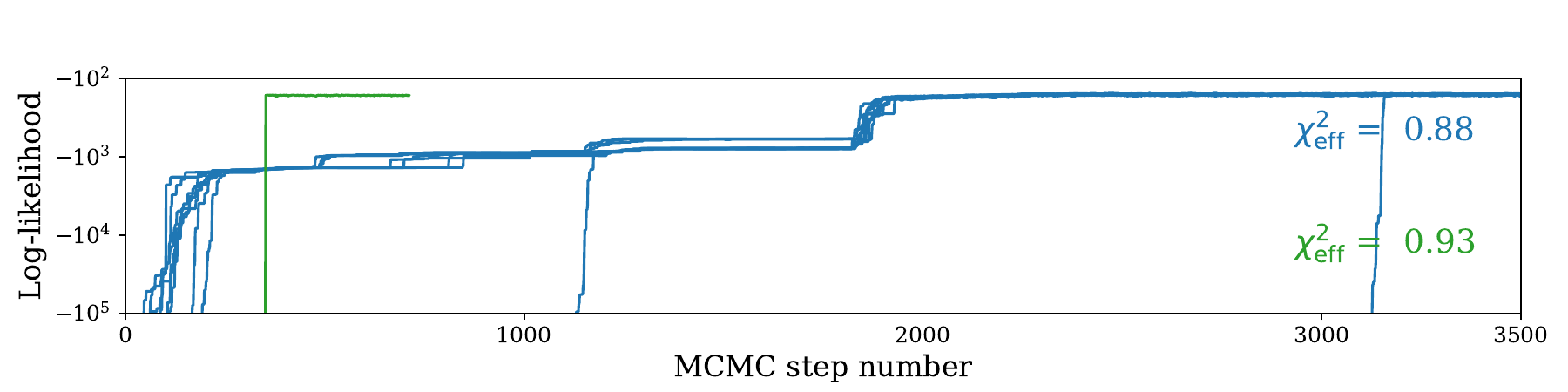}
    \caption{Trace of the chains fitting all 4 parameters for different samplers (AI: affine invariant, AFS: automated factor slice). The chains were initialized in the same way as for model A. AI was run with 40 cores and the computational efficiency was 1 step per core per hour. AFS was run with 12 cores and the efficiency was the same, 1 step per core per hour, with a much faster convergence though.}
    \label{fig:P4_samps}
\end{figure*}

Fig. \ref{fig:P4_samps} shows the fitting procedure for all parameters using AI and AFS. For initial values we chose the same as in the main results, that is $i_0=90$, $M_{\rm unit,0}=8\times10^{18}$, $R_{\rm high, 0}=30$, $\rm PA_0$=0. It is evident that AFS converges faster in this example. The fact that some chains from AI get stuck for a significant amount of time can be a problem for more advance fitting procedures. This is a well known problem in certain fits with AI, reported already in \cite{Foreman-Mackey2013PASP125.306}.

In Fig.~\ref{fig:Mu_ranges} we show the chain evolution for the AI sampler for different initial range, using the same initial value. The convergence is faster with a wide range if the initial value is far from the truth, but with better informed prior a smaller range is to be preferred, as stated in \cite{Foreman-Mackey2013PASP125.306}. 

The combination of these tests, taking into account the computational time, makes AI our sampler of choice for this work. Furthermore we choose to initialize our chains near the middle of the desired area of exploration with an initial range that covers it all to be as agnostic as possible. In the future a hybrid approach, using AI for a wide survey and AFS for a more detailed exploration, with priors obtained from AI, could prove optimal. 

\bsp	% typesetting comment
\label{lastpage}
\end{document}